\documentclass[showpacs,amsmath,amssymb,twocolumn,prx,superscriptaddress,notitlepage,aps]{revtex4-2}

\usepackage{qcircuit}
\usepackage[dvips]{graphicx}
\usepackage{amsmath,amssymb,amsthm,mathrsfs,amsfonts,dsfont}
\usepackage{subfigure, epsfig}
\usepackage{braket}
\usepackage{hyperref}
\usepackage{bm}
\usepackage{enumerate}
\usepackage{color}
\usepackage{graphicx}

\usepackage{amsthm}

\newtheorem{result}{Result}
\newtheorem{lemma}{Lemma}
\newtheorem{theorem}{Theorem}
\newtheorem{remark}{Remark}
\newtheorem{definition}{Definition}

\newcommand{\tr}{\mathrm{Tr}}

\newcommand{\cf}{\mathbf{F}_C}
\newcommand{\qf}{\mathbf{F}_Q}
\newcommand{\matra}{\mathbf{A}}
\newcommand{\matrm}{\mathbf{M}}
\newcommand{\f}{\lambda}
\newcommand{\spl}{Appendix}

\begin{document}


\title{Quantum natural gradient generalised to noisy and non-unitary circuits}


\author{B\'alint Koczor}
\email{balint.koczor@materials.ox.ac.uk}

\affiliation{Department of Materials, University of Oxford, Parks Road, Oxford OX1 3PH, United Kingdom}
\affiliation{Quantum Motion, 9 Sterling Way, London N7 9HJ, United Kingdom}

\author{Simon C. Benjamin}
\email{simon.benjamin@materials.ox.ac.uk}
\affiliation{Department of Materials, University of Oxford, Parks Road, Oxford OX1 3PH, United Kingdom}
\affiliation{Quantum Motion, 9 Sterling Way, London N7 9HJ, United Kingdom}


\begin{abstract}
Variational quantum algorithms are promising tools whose efficacy depends on their optimisation method. For noise-free unitary circuits, the quantum generalisation of natural gradient descent has been introduced and shown to be equivalent to imaginary time evolution: the approach is effective due to a metric tensor reconciling the classical parameter space to the device's Hilbert space. Here we generalise quantum natural gradient to consider arbitrary quantum states (both mixed and pure) via completely positive maps; thus our circuits can incorporate both imperfect unitary gates and fundamentally non-unitary operations such as measurements. We employ the quantum Fisher information (QFI) as the core metric in the space of density operators. A modification of the Error Suppression by Derangements (ESD) and Virtual Distillation (VD) techniques enables an accurate and experimentally-efficient approximation of the QFI via the Hilbert-Schmidt metric tensor using prior results on the dominant eigenvector of noisy quantum states. Our rigorous proof also establishes the fundamental observation that the geometry of typical noisy quantum states is (approximately) identical in either the Hilbert-Schmidt metric or as characterised by the QFI.
In numerical simulations of noisy quantum circuits we demonstrate the practicality of our approach  and confirm it can significantly outperform other variational techniques. 
\end{abstract}

\maketitle

\section{Introduction}

Variational techniques are ubiquitous in physics and mathematics. More specifically, variational algorithms involving the incremental update of parameters that describe a many-body quantum state have been widely used for decades~\cite{BALIAN198829,RevModPhys.71.463,PhysRevLett.107.070601,SHI2018245,vanderstraeten2018tangent}. The technique involves using a tractable set of parameters to describe a quantum state in an exponentially larger Hilbert space, and therefore relies on an understanding that the states of importance (e.g. the low-energy states of some Hamiltonian) lie within a relatively small subspace. 

With the rise of quantum computers as realistic technologies, naturally attention has been given to the question of how such a machine could perform as a variational tool~\cite{farhi2014quantum,peruzzo2014variational,wang2015quantum,PRXH2,PhysRevA.95.020501,mcclean2016theory,PhysRevLett.118.100503,Li2017,PhysRevX.8.011021,Santagatieaap9646,kandala2017hardware,kandala2018extending,PhysRevX.8.031022,romero2017strategies,higgott2018variational,mcclean2017hybrid,colless2017robust,kokail2018self}.
The resulting model is hybrid with an iterative loop: a classical machine determines how to update the parameters describing a quantum state (the `ansatz state') while the quantum coprocessor generates and characterises that state (using an `ansatz circuit'). This paradigm is of particular interest in the context of noisy, intermediate-scale quantum devices (NISQ devices)~\cite{preskill2018quantum}, because quite complex ansatz states can be prepared with shallow  circuits~\cite{kassal2011simulating,C2CP23700H,whaley2014quantum,ourReview}, thus raising the possibility that resource-intensive quantum fault tolerance methods might not be needed. 

The most well-studied application is the variational quantum eigensolver (VQE), where one seeks a final circuit configuration that minimises a cost function -- normally the energy of some system of interest. For the optimisation of the classical parameters, one might employ any one of a range of methods: for example a direct search such as Nelder–Mead (demonstrated experimentally in 2014~\cite{peruzzo2014variational}), or a systematic scan if the number of parameters is small~\cite{PRXH2}, or direct gradient descent (see e.g. ref.~\cite{Rebentrost_2019}). 

An imaginary time principle can also be used to govern the parameter evolution, so that the ansatz state follows (as closely as possible) the trajectory $e^{-t \mathcal{H}} |\psi_0\rangle$~\cite{samimagtime}. The approach was found to outperform others in accuracy and convergence speed according to numerical simulations~\cite{samimagtime}, and was subsequently demonstrated experimentally~\cite{imagTimeDemo2019}. Deriving the parameter evolution, which proceeds from McLachlan's variational principle as in real-time quantum evolution~\cite{Li2017}, introduces an important feature: a matrix object that characterises the sensitivity of the ansatz state to changes in each possible pair of parameters, but without reference to the cost function. 

Indeed this matrix has a crucial role in enabling the high performance of the technique,
however, an elucidation of its deeper meaning was reached in relation to a concept called {\it natural gradient}~\cite{quantumnatgrad,yamamoto2019natural,amari1997neural,goodfellow2016deep,amari2000adaptive}.
Natural gradient accounts for the non-trivial relationship between a translation in parameter space, and the corresponding translation in the problem space (for our case, the Hilbert space of the quantum processor).
Ref.~\cite{quantumnatgrad} illustrated the approach
by the example of a state-vector that is isomorphic to a classical probability distribution
$p( n |\underline{\theta} )$ (i.e., the state contains no phase information):
In this case  both imaginary time evolution and natural gradient
result in the update rule of variational parameters $\underline{\theta}(t)$ indexed by $t$ as
\begin{equation} \label{naturalgradEvo}
	\underline{\theta}(t{+}1) = \underline{\theta}(t) - \kappa \, [\cf]^{-1} \underline{g}.
\end{equation} 
Here the classical Fisher information matrix $\cf$ is a metric tensor
that is related to the previous classical probability distribution via
\begin{equation}\label{cfisher}
	[\cf]_{kl}  =  \sum_n p( n |\underline{\theta} )\frac{\partial^2 \, \mathrm{ln} \, [p( n |\underline{\theta} )]}{\partial \theta_k \, \partial \theta_l} 
\end{equation}
and its inverse corrects the gradient vector by accounting for the co-dependent and
non-uniform effect of the parameters on $p( n |\underline{\theta} )$.

\begin{figure*}[tb]
	\begin{centering}
		\includegraphics[width=0.95\textwidth]{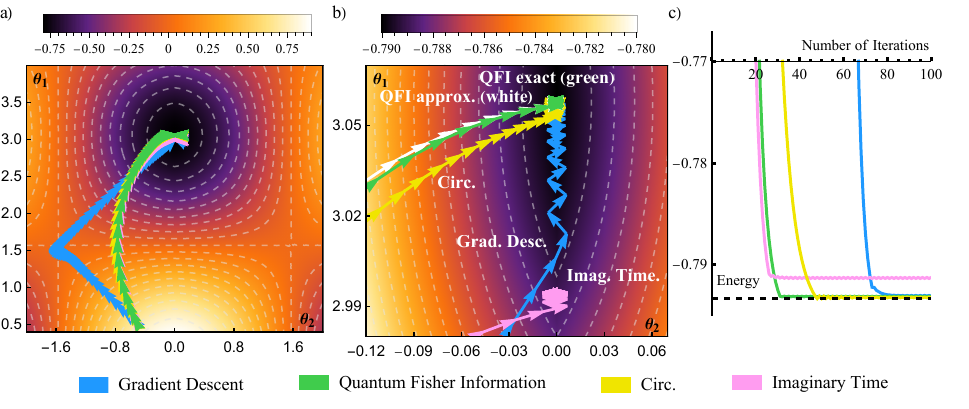}
		\caption{
			Different strategies for finding ground states using noisy quantum circuits --
			our natural gradient approach using the quantum Fisher information matrix as a metric tensor
			(in the space of \emph{mixed} states) outperforms other methods
			both in terms of accuracy and number of steps.			
			a-b) Energy landscape of a 2-qubit Hamiltonian
			$\mathcal{H} = Z_1 + 0.1 X_1 X_2$
			as a function of the two parameters $(\theta_1,\theta_2)$ of a noisy quantum circuit and colours
			represent the energy $E= \tr[\rho (\theta_1,\theta_2) \mathcal{H}]$.
			Gradient descent evolution (blue) may get trapped a)
			and converges slowly in the close vicinity of the optimum b).
			Imaginary time evolution (of mixed states as in Appendix~\ref{densityimag}) and natural gradient evolution from
			Result~\ref{result1} (pink vs. green) reach the optimum rapidly a), but
			the former does not converge to the true optimum b).
			The main practical result of the present work is a noise-aware approximation of the QFI matrix (white) that
			fits very well with exponentially powerful error mitigation techniques~\cite{koczor2020exponential,huggins2020virtual}.	
			Our approach (white) impressively well approximates the quantum Fisher information, significantly
			better than emulating pure-state imaginary time evolution using noisy, experimental
			Hadamard-test circuits~\cite{samimagtime}(yellow) as expected from our rigorous proofs.
			\label{introfig}
		}
	\end{centering}
\end{figure*}

While the approach was thought to be more expensive than gradient descent due to the need for
	evaluating a matrix object, ref.~\cite{van2020measurement} rigorously proved that the measurement overhead of the
natural gradient approach is asymptomatically negligible in many practical scenarios. Given 
the faster convergence rate of the natural gradient approach it is expected to reach the optimum with fewer quantum
resources and is therefore provably superior to simple gradient descent.

In the present paper we propose a novel quantum variational optimisation
method which is directly analogous to Eq.~\eqref{naturalgradEvo}
but can be applied to arbitrary mixed or pure states.
Our approach thus generalises and unifies previously obtained state-vector evolutions
to the non-trivial and most general case of density operators
(both mixed and pure states). This is particularly relevant when considering
the effect of imperfections on variational quantum circuits
as illustrated in Fig.~\ref{introfig} but allows for
additional degrees of freedom in the optimisation, such as
non-unitary transformations including measurements.

The most important practical result of the present work is an
approach that can be used in tandem with exponentially powerful error mitigation
techniques~\cite{koczor2020exponential, huggins2020virtual}
and can be applied to optimising noisy quantum circuits.
This results in a `noise-aware' natural gradient evolution whereby
the matrix object is determined efficiently as an approximation to
the quantum Fisher information and gradients are determined using error mitigation
techniques.
We finally present rigorous proofs on the error robustness of our approach
building on prior theoretical results on the dominant eigenvector of a noisy quantum state~\cite{koczor2021dominant}.

This manuscript is organised as follows. In the rest of this introduction we recapitulate prior work
on variational quantum optimisation using idealised quantum circuits in
Sec.~\ref{recap} and then discuss in Sec.~\ref{circuitSec} how noise affects
a variational quantum circuit  --  an experienced reader can skip these sections.
In Sec.~\ref{sec:application} we motivate our main results 
by illustrating our approach on practical examples and we demonstrate that it
significantly outperforms other techniques when quantum circuits are noisy.
We then introduce our general natural gradient evolution in Sec.~\ref{mainres} and relevant
notions in the context of quantum information geometry.
We finally state our main theoretical results in Sec~\ref{approxsec}
and establish that accurate and efficient optimisation should be possible
for a comprehensive range of NISQ-era scenarios.

\subsection{Optimising idealised variational quantum circuits\label{recap}}

An idealised variational quantum circuit is modelled as a unitary transformation
$ | \psi(\underline{\theta}) \rangle = U_c(\underline{\theta}) \, |00\dots0 \rangle$
applied to the reference state $|00\dots0 \rangle$ which is usually chosen
as the computational zero state. The unitary operator $U_c(\underline{\theta})$
represents the entire quantum circuit and depends on a set of gate parameters
$\underline{\theta} \in \mathbb{R}^\nu$. As elucidated by the circuit model,
it decomposes into a series of individual gates (typically acting on a small subset of
the system, i.e.,  on one or two qubits)
$$ U_c(\underline{\theta}) = U_\nu(\theta_\nu) \dots  U_2(\theta_2) U_1(\theta_1),$$
each of which depends on a parameter $\theta_i$ with $i\in\{1, 2, \dots \nu \}$.
It is typically the aim of a variational algorithm to find the minimum or maximum
of an expectation value
$E(\underline{\theta}) := \tr[\rho(\theta) \mathcal{H}] = \langle \psi(\underline{\theta}) | \mathcal{H}| \psi(\underline{\theta}) \rangle $
over the parameters where the observable $\mathcal{H}$ is a Hermitian operator,
typically  the Hamiltonian of a simulated physical system. A hybrid approach assumes
that a quantum processor can efficiently estimate the expectation
value $E(\underline{\theta})$ for a set of parameters and these parameters
are optimised externally according to an update rule calculated by a classical computer. 

As outlined in the introduction, numerous optimisation methods have been proposed for finding 
parameters that minimise this expectation value and indeed some have been demonstrated experimentally.
Imaginary time evolution~\cite{samimagtime,xiaotheory} and
the natural gradient evolution~\cite{quantumnatgrad}
optimise the parameters $\underline{\theta}(t)$ iteratively in steps $\Delta t$ as
\begin{equation}\label{imagtimeUpdate}
	\underline{\theta}(t{+}1) = \underline{\theta}(t) - \Delta t \, \matra^{-1} \underline{g}.
\end{equation}
Here the inverse of the matrix $\matra:= \matra(\underline{\theta})$ is applied to
the gradient of the expectation value
$g_k =  \partial_k E(\underline{\theta})$ and we will consistently use
the short-hand notation $\partial_k := \tfrac{\partial}{\partial \theta_k}$.
The matrix object $\matra$ in the update rule has been related to the well-known Fubini-Study metric tensor~\cite{study1905kurzeste,fubini1904sulle,wilczek1989geometric,hackl2020geometry}
that is defined in complex projective spaces $\mathbf{CP}^n$ as the real part of the so-called quantum geometric tensor $G_{kl}$ and whose
matrix elements are given by the state-vector overlaps \cite{study1905kurzeste,fubini1904sulle,wilczek1989geometric,hackl2020geometry}
\begin{equation}\label{amatrix_def}
	[\matra]_{kl} := \mathrm{Re} [  G_{kl} ]  = \mathrm{Re}[\langle \partial_k \psi | \partial_l \psi \rangle 
	- \langle \partial_k \psi | \psi \rangle \langle  \psi | \partial_l \psi \rangle].
\end{equation}
The second term above was dropped in ref.~\cite{samimagtime}
and global phase gauge was enforced by the introduction of a quantum gate
(see, e.g., Eq.~(4) in~\cite{cheng2010quantum} for an intuitive explanation)~\footnote{
			The second term in Eq.~\eqref{amatrix_def} vanishes when the global phase evolution
			under variations of the parameters is zero. For example, ref.~\cite{samimagtime} dropped this term and
			ensured zero global phase evolution through the introduction of
			a single redundant gate. However, generally one needs to
			take into account global phase evolution and McLachlan’s variational principle
			was therefore modified in ref.~\cite{xiaotheory} where Eq.~(14) includes the second term due
			to global phase evolution.
			Further details are also discussed in Sec.~4.1.3 of~\cite{hackl2020geometry}
}.
In the following we will consistently use the short-hand notation throughout this paper
$\partial_k \psi := \tfrac{\partial \psi(\underline{\theta})}{\partial \theta_k} $.

The update rule in Eq.~\eqref{imagtimeUpdate} was
originally derived in refs. \cite{samimagtime,xiaotheory} to simulate the imaginary time evolution of 
a pure state vector
\begin{equation}\label{imagevo}
	|\psi(t)\rangle = \frac{e^{-t \mathcal{H}} |\psi_0\rangle } { \langle \psi_0 | e^{-2 t \mathcal{H}} |\psi_0 \rangle },
\end{equation}
using a variational quantum circuit that can efficiently estimate
both $\matra$ and $\underline{g}$. The exact evolution in Eq.~\eqref{imagevo} converges to the ground state of the
system for $t \rightarrow \infty$ if $|\psi_0 \rangle$
has a non-zero overlap with the ground state -- hence it is guaranteed to avoid local minima given a sufficiently
deep ansatz circuit.
We also note that quantum natural gradient and imaginary time evolution are not Hessian-based Newton optimisations
of the energy surface as discussed in \cite{quantumnatgrad}:
While the quantum Fisher information can be related to a Hessian matrix in special cases,
e.g., the classical Fisher information is the Hessian of a KL-divergence,
it is not the Hessian of the energy surface as it makes no reference to the cost function.

The above discussed methods were derived for idealised \emph{perfect} quantum circuits and pure states,
and naively applying these equations might be suboptimal when noise is present. 
Fig.~\ref{introfig} (white) illustrates that our noise-aware protocol gives a remarkably accurate approximation of
 the quantum Fisher information, significantly better than emulating pure-state imaginary time evolution
with experimental, noisy Hadamard-test circuits of ref.~\cite{samimagtime} (yellow).

\subsection{Variational algorithms with non-unitary circuits\label{circuitSec}}

We first generalise the previously introduced idealised unitary circuit model
to the more realistic case taking experimental imperfections of the variational
circuit into account. We describe this variational circuit as a
completely positive map of density matrices as 
$\rho(\underline{\theta})  = \Phi(\underline{\theta}) \, \rho_0$ that
depends on the parameters $\underline{\theta} \in \mathbb{R}^\nu$ and
$\rho_0$ is usually the computational zero state. Here $\Phi(\underline{\theta})$
is the superoperator that represents the realistic quantum circuit.
This quantum circuit only \emph{approximately}
decomposes into a sequence of imperfect gate operations $\Phi_k(\theta_k)$  as
$$ \Phi(\underline{\theta}) \approx \Phi_\nu(\theta_\nu) \dots \Phi_2(\theta_2) \Phi_1(\theta_1), $$
due to possible correlated noise.

Our approach is, however, not restricted to imperfect quantum
circuits. The only assumption we make about the
mapping $\Phi(\underline{\theta})$ is that it is
continuous in each of the parameters $\theta_k$ such that differentials $\partial_k \rho(\underline{\theta})$ of the density matrix as
$[\partial_k \Phi(\underline{\theta})] \, \rho_0$
exist for any state.
This is naturally the case for quantum circuits
undergoing (possibly time-dependent) Markovian
decoherence \cite{infdivchannel1} but more general
mappings can satisfy this condition too. These
include, e.g., allowing measurements in the circuit
independently of the parametrisation or decoherence whose length
depends on a parameter. Refer to Sec.~\ref{obj-func-sec} for more discussion. We further note that our characterisation also naturally
generalises to infinite-dimensional Hilbert spaces
as continuous-variable systems by using
continuous maps $\Phi(\underline{\theta})$ over trace-class
operators $\rho$ \, \cite{ReedSimon1}.

We note here that a generalisation to mixed quantum states of the previously discussed imaginary time evolution
was proposed in \cite{xiaotheory}, but this approach does not necessarily converge to
the true optimum as illustrated in Fig.~\ref{introfig} (pink). Refer to Appendix~\ref{densityimag} for more details.
In the following we aim to develop an alternative approach that
does not rely on the simulation of imaginary time evolution
(yet reduces to the unitary variant in case of pure states) and finds the true optimum
as illustrated in Fig.~\ref{introfig} (green).

\begin{figure*}[tb]
	\begin{centering}
		\includegraphics[width=0.95\textwidth]{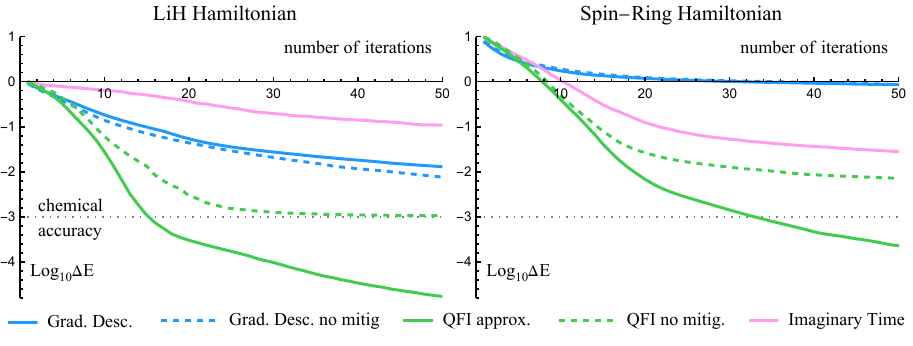}
		\caption{
				Average energy distance from the ground state $\Delta E$ of two $N=6$-qubit Hamiltonians achieved in $30$ runs with various optimisation techniques.
				Two-qubit gates undergo errors with a probability $10^{-3}$ while
				gradients were determined with (blue and green solid lines) and without (blue and green dashed lines) ESD/VD error mitigation.
				Natural gradient (green lines) using our noise-aware approximation of the QFI matrix
				requires the same resources as the ESD/VD error mitigation techniques. 
				(pink) imaginary time evolution (of mixed states from Appendix~\ref{densityimag}).
				All energies $\Delta E$ were computed as if we had a perfect noiseless circuits 
				at parameters obtained from the noisy optimisation. 
			\label{fig:vqe_plots}
		}
	\end{centering}
\end{figure*}

\section{Application Examples\label{sec:application}}

Before deriving our main results, we first provide two examples
of how our approach can be used in practice.
In particular, in Sec.~\ref{approxsec} we will derive our noise-aware natural gradient approach whereby
the quantum Fisher information is
well approximated by the Hilbert-Schmidt metric tensor that we can efficiently determine. As we explain later, our approach requires
the exact same resources as the powerful Error Suppression by Derangements (ESD~\cite{koczor2020exponential}) and Virtual Distillation (VD~\cite{huggins2020virtual}) error mitigation techniques:
we require two copies of the noisy computational state that we use to effectively verify each other
via a so-called derangement circuit that consists of controlled-SWAP operations. The approach has been shown to 
be very NISQ-friendly; It also lends itself for a multicore architecture whereby macroscopically separate
quantum cores independently prepare the two copies of the noisy quantum state  which are then entangled via noisy quantum links;
state-of-the art silicon technology has been shown to be a suitable platform~\cite{jnane2022multicore}.

In the following we demonstrate how our approach can be used in tandem with ESD/VD for preparing ground states
of Hamiltonian models using variational quantum circuits assuming typical hardware-noise characteristics.
First, we consider a second-quantised LiH Hamiltonian~\cite{ourReview} of $N=6$ qubits
and in Fig.~\ref{fig:vqe_plots}(left) we plot the average achieved distance $\Delta E$ from the ground state over $30$ simulations
each started from random initial parameters around the Hartree-Fock solution.
Our noise-aware natural gradient approach (solid green line) significantly outperforms all other techniques by orders of magnitude
and comfortably achieves chemical accuracy as $\Delta E \leq 10^{-3}$.
However, when not using ESD/VD error mitigation for the gradient estimation (green dashed) the optimisation approaches a noise floor
above chemical precision.

Second, we consider a spin-ring Hamiltonian of $N=6$ qubits
with a constant coupling $J=1$ and uniformly randomly generated on-site interaction strengths $\omega_k \in [-1,1]$ as 
\begin{equation}\label{spin_ring}
	\mathcal{H} =  \sum_{k \in \text{ring}(N)} \omega_k Z_k  + J \, \vec{\sigma}_k \cdot \vec{\sigma}_{k+1}.
\end{equation}
This Hamiltonian is relevant in the context of manybody localisation and in demonstrating practical near-term quantum advantage~\cite{nandkishoreManyBodyLocalizationThermalization2015,childsFirstQuantumSimulation2018}. We show the average of 30 optimisations started randomly in parameter space in Fig.~\ref{fig:vqe_plots}(right).
As in the previous example, our noise-aware natural gradient approach (solid green line) significantly outperforms all other techniques
when the gradient vector is determined using error mitigation. Furthermore, in this example it is even more pronounced
how conventional gradient descent (blue lines) is more vulnerable to getting stuck in local optima (significantly higher average $\Delta E$).

These examples yield the following observations.
a) our noise-aware 
protocol approximates the QFI matrix very well due to the suppression of hardware errors 
as expected from
our rigorous error bounds in Result~\ref{result3};
b)  given we need to estimate the gradient vector from a noisy experiment,
our natural gradient approach is still limited by noise in the gradient vector
--
this is nicely illustrated in Fig.~\ref{fig:vqe_plots}(green dashed lines)
where the evolution hits a noise floor and cannot reach chemical precision without error mitigation;
c) in Fig.~\ref{fig:vqe_plots} (green solid lines) we use ESD/VD error mitigation
to estimate the gradient vector and find a very good approximation of the
ground state, with energy deviation $\Delta E \ll 10^{-3}$;
d) Our approach performs impressively well on both problem Hamiltonians: such 
robustness is expected given the QFI matrix is completely independent of the
problem Hamiltonian.

Such a low achieved energy with a noisy optimiser is impressive
given our error mitigation approach is limited by a coherent mismatch~\cite{koczor2021dominant}.
Whereas it was already expected in refs.~\cite{koczor2020exponential, huggins2020virtual}
that variational quantum algorithms by construction minimise coherent errors, Fig.~\ref{fig:vqe_plots} clearly demonstrates
that indeed a variational optimisation minimises the effect of this coherent mismatch within the variational subspace
and thus boosts the performance of ESD/VD.
Of course, achieving extremely precise ground state energies
is hindered in practice by shot noise, i.e., by finite circuit repetition;
While measurement costs are comprehensively explored in ref.~\cite{van2020measurement},
here we assumed a sufficiently large number of samples
to suppress the impact of shot noise
as our aim is to focus on fundamental limitations posed by hardware noise.

Nevertheless, we can nicely illustrate on our LiH example
why the sampling cost of the metric tensor may become negligible
asymptotically for an increasing number of qubits $N$
or number of iterations $t$~\cite{van2020measurement}:
the Hamiltonian may consist
of a large number $r_H$ of Pauli terms and thus determining the gradient vector
may need a number of samples $\propto r_h \nu$.
In contrast, determining the QFI matrix is independent of the Hamiltonian and
requires a number of samples $\propto \nu^2$. The key observation is that the number of
parameters in a \emph{shallow quantum circuit} can only grow mildly, e.g., as $\propto N \mathrm{polylog}(N)$ whereas
the number of Hamiltonian terms may grow much faster, e.g., as $r_H = \mathcal{O}(N^4)$ in case of molecular systems.
Similarly, as approaching
the optimum for an increasing number of iterations makes gradients vanish,
determining their entries to a comparable precision to that of the metric tensor becomes increasingly expensive.

We estimate explicit sampling costs for our LiH problem in Appendix~\ref{app:figvqeplots} and find
that determining the gradient vector to a typical practical level of precision $\epsilon = 0.05$ requires
about $2.6 \times 10^6$ samples whereas determining all QFI matrix entries to the same precision requires
fewer, about $2.1 \times 10^6$ samples. Furthermore, as we decrease the sampling precision of the gradient vector, e.g.,
to $\epsilon = 0.01$ as required for reaching chemical precision, the cost of properly sampling the gradient vector dominates
over the QFI sampling cost by an order of magnitude ($\approx 6.5 \times 10^7$).
Assuming sub-millisecond circuit execution time, as is realistic for solid state (superconducting or silicon spin) devices, then collecting samples for a single iteration may take several minutes; however this is perfectly parallelizable over multiple quantum processors~\cite{PhysRevLett.129.010502, caiResourceEstimationQuantum2020, jnane2022multicore}.

\section{Quantum Fisher information and natural gradient\label{mainres}}
Before deriving our main results, we briefly recall basic notions
related to the quantum Fisher information which is a concept extensively
used in the field of quantum metrology for determining
the metrological usefulness of a quantum state,
refer to, e.g., \cite{review,giovannetti11,koczor2019variational} for more details.
While equivalent expressions have already been proposed in the context
of pure quantum states, here we introduce the quantum Fisher information
in the  context of general variational quantum circuits as a measure that
quantifies how much and in what way changing parameters in a quantum
circuit affects the underlying (in general, mixed) quantum state.

\subsection{Sensitivity to parameters via the quantum Fisher information\label{qfishersec}}
Since the quantum Fisher information is defined in terms of notions adapted from classical
probability theory \cite{qfi1} (in contrast to the Fubini-Study metric tensor which is defined in complex projective spaces), we introduce and illustrate these core concepts first.
Assume now for simplicity a one-parameter quantum circuit as a one-parameter mapping
$\rho_\theta = \Phi(\theta) \rho$ acting on a reference state $\rho$
and let us consider the resulting continuous family of quantum states $\rho_\theta$.
Here $\Phi(\theta)$ can be, e.g., an imperfect M\o lmer-S\o rensen gate.
The quantum Fisher information is a generalisation of the classical Fisher information
and quantifies how different a state $\rho_\theta$ 
becomes under an infinitesimal variation $\rho_{\theta+\delta \theta}$ of
this gate parameter $\theta$.

The aforementioned parametrised state produces a family of probability distributions
$$p( n |\theta ) = \tr[\rho_\theta \, |b_n \rangle \langle b_n |] =  \langle b_n |  \rho_\theta  |b_n \rangle $$
when measured in a basis $\{ |b_n \rangle \}$, where
$n=\{1,2,\dots, d\}$, $\sum_n |b_n \rangle \langle b_n | = \mathrm{Id_d}$ and $d$ is the dimensionality of the system.
The classical Fisher information quantifies how different these \emph{probability
	distributions} become under a variation of $\theta$
\begin{equation*}
	F_c(\{ |b_n \rangle \})  =  \sum_n p( n |\theta ) \bigg(\frac{\partial\mathrm{ln}    ~ p( n |\theta )}{\partial \theta} \bigg)^2,
\end{equation*}
and depends on the choice of measurement basis $\{ |b_n \rangle \}$.
For example, if the parameter $\theta$ corresponds to a z-rotation
$U_z(\theta):=\exp(-i\theta \sigma_z/2)$ of a qubit state as $\rho_\theta=U_z(\theta) \rho [U_z(\theta)]^\dagger$,
then measuring in the computational basis results in, e.g.,
$\partial_\theta \langle 0 |\rho_\theta | 0 \rangle = 0$ and therefore the Fisher information is
$F_c(\{ |0 \rangle , |1 \rangle\}) = 0$. However, 
measuring in the $|\pm\rangle$ basis results in $F_c(\{ |+ \rangle , |- \rangle\}) \geq 0$.

In the present single-parameter case, the quantum Fisher information is the maximum of $F_c(\{ |b_n \rangle \}) $ when optimised
over all possible measurement basis sets as $\{ |b_n \rangle \}$ (including generalised POVM measurements) \cite{qfi1}.
This quantum Fisher information, which depends on the state and the current parameter values $F_Q = F_Q(\rho_\theta) \geq 0$,
\emph{is defined} \cite{qfi1,review} as the expectation value $F_Q(\rho_\theta) := \tr[\rho_\theta L_\theta^2]$.
Here the Hermitian symmetric logarithmic derivative $L_\theta$ contains the most sensitive
measurement basis (with respect to a variation of $\theta$) as eigenvectors and is defined via
\begin{equation} \label{symmlogder}
	\frac{\partial \rho_\theta}{\partial \theta} =: \frac{1}{2}(L_\theta \rho_\theta +\rho_\theta L_\theta).
\end{equation}
Decomposing a  density matrix into $\rho_\theta = \sum_{n} p_n |\psi_n \rangle \langle \psi_n | $
projectors onto its eigenstates $|\psi_n \rangle$ with $p_n>0$
allows for explicitly computing matrix elements of the symmetric logarithmic derivative as
\begin{equation*}\label{ll}
	\langle \psi_k | L_\theta | \psi_l \rangle = \frac{2}{p_k + p_l} \langle \psi_k | \frac{\partial \rho_\theta}{\partial \theta} | \psi_l \rangle.
\end{equation*}

\subsection{Quantum Fisher information matrix\label{qfishermatsec}}
Let us now consider the matrix form of the quantum Fisher
information which is a quantum generalisation of the
classical Fisher information matrix from
Eqs.~(\ref{naturalgradEvo}-\ref{cfisher}).
We will now consider parametrised quantum circuits (or more generally continuous mappings) from Sec.~\ref{circuitSec}
that span a continuous family of  density matrices as $\rho(\underline{\theta})$.
As in Eq.~\eqref{symmlogder},
the partial derivative of $\rho(\underline{\theta})$ with respect
to an individual parameter $\theta_k$ defines the symmetric logarithmic derivative
\begin{equation} \label{symmlogdermulti}
	\partial_k \rho(\underline{\theta})  =: \frac{1}{2}(L_k \rho(\underline{\theta}) +\rho(\underline{\theta}) L_k),
\end{equation}
and eigenvectors of $L_k$ are the most sensitive measurement bases to detect variations in $\theta_k$.
Entries of the quantum Fisher information matrix $[\qf]_{kl}=[\qf \{\rho(\underline{\theta})\}]_{kl}$,
which depend on both the state and the current parameter values,
are \emph{defined as} the expectation values
\begin{equation} \label{qfi-mat-def}
	[\qf]_{kl}=[\qf]_{lk} := \tfrac{1}{2} \tr[\rho(\underline{\theta}) \, (L_k L_l + L_l L_k) ]
\end{equation}
of symmetric logarithmic derivatives. 
Diagonal entries of $\qf$ correspond to
the scalar quantum Fisher information 
$\tr[\rho_\theta L_k^2]$ and quantify the sensitivity of a
quantum state with respect to individual parameters $\theta_k$.
Off-diagonal entries account for the co-dependence of parameters.
Note that the above definition of a metric tensor is fundamentally different to that
of the Fubini-Study metric tensor \cite{hackl2020geometry}.

\subsection{Geometry of quantum states and metric tensors\label{geometry_sec}}

It is well known that the imaginary time evolution of pure states
(as obtained via the McLachlan principle) defines a Riemannian gradient descent, i.e.,
a gradient descent that takes into account the Riemannian geometry of pure states
(refer to Proposition~11 in \cite{hackl2020geometry} for more details).
Indeed, the pure-state natural gradient approach \cite{quantumnatgrad} is
equivalent to the imaginary time principle~\cite{samimagtime}, and it is commonly known that
the Fubini-Study metric tensor $\matra$ from Eq.~\eqref{amatrix_def} is the obvious unique metric tensor in the
space of pure states \cite{study1905kurzeste,fubini1904sulle,wilczek1989geometric,hackl2020geometry}.

Here we consider a Riemannian gradient descent for arbitrary quantum circuits;
In contrast to unitary circuits for state vectors, this problem becomes non-trivial for the case of mixed quantum states
as one obtains a family of an infinite number of contractive Riemannian
metric tensors \cite{qfi1,morozova1991markov, petz_geometry,petz_book_chapter}.
Among these, the most natural choice for our purposes is the quantum Fisher information $\qf$
for the following 4 reasons (refer also to Theorem~3.4 in \cite{petz_book_chapter} for more discussion).

a) $\qf$ is a Riemannian metric tensor and the
suitably defined distance~\cite{qfi1,morozova1991markov, petz_geometry,petz_book_chapter}
in state space $\mathrm{d}s^2$
can be expressed in terms of the usual infinitesimal (first order) variations of the ansatz parameters
\begin{equation}\label{metric_def}
	\mathrm{d}s^2 = \sum_{k,l =1}^\nu \tfrac{1}{4} [\qf]_{kl} \, \mathrm{d} \theta_k \, \mathrm{d} \theta_l.
\end{equation}

b) If the rank of the density matrix does not change
under the variation of the parameters~\cite{qfi_discont,qfi1} then $\qf$ reduces to the so-called Bures metric
tensor as
$[\qf]_{kl} = 4 g_{kl}$, refer to \cite{bures1969extension,braunstein94,petz_geometry,petz_book_chapter,qfi1}.
In this case the distance $\mathrm{d}s^2 $ in Eq.~\eqref{metric_def} can be specified
in terms of the fidelity \cite{braunstein94, uhlmann1976transition}
between the two quantum states $\rho_1:=\rho(\underline{\theta})$ and $\rho_2:=\rho(\underline{\theta} + \mathrm{d}\underline{\theta})$
as
\begin{equation*}
	\mathrm{d}s^2 = 2 (1-\sqrt{\mathrm{Fid}(\rho_1, \rho_2)}) = 2(1-\tr[  \sqrt{   \sqrt{\rho_1} \rho_2  \sqrt{\rho_1}   }  ]).
\end{equation*}

c) The $\qf$ naturally reduces to the aforementioned
Fubini-Study metric tensor in the limiting case of  pure states
as $[\qf]_{kl} = 4 \matra_{kl}$, as shown by Petz et al. \cite{petz_geometry,petz_book_chapter},
refer also to \cite{qfi1,morozova1991markov}. 
The distance metric $\mathrm{d}s^2 $ in Eq.~\eqref{metric_def} is then specified
as $\mathrm{d}s^2 = \arccos |  \langle \psi_1 | \psi_2 \rangle | $
in terms of the pure states $\psi_1:=\psi(\underline{\theta})$ and $\psi_2:=\psi(\underline{\theta} + \mathrm{d}\underline{\theta})$.
This ensures us that our approach naturally reduces to the pure-state 
techniques introduced in Sec.~\ref{recap} which we further detail in Sec.~\ref{pure-state-sec}.

d) In general, the QFI is the maximum
of the classical Fisher information (from Eq.~\eqref{class-fisher}) when optimised over all measurement bases.
The distance metric in Eq.~\eqref{metric_def} is related to the so-called Bhattacharyya coefficient
which characterises the similarity via
$\mathrm{d}s^2  = 2-2\sum_n \sqrt{p( n |\underline{\theta} ) p( n |\underline{\theta} {+} \mathrm{d}\underline{\theta})}$
of the respective, maximally sensitive measurement probabilities \cite{braunstein94,petz_geometry,petz_book_chapter,qfi1}.
It follows that when the quantum state is isomorphic to a classical probability
distribution (in the computational basis), then the $\qf$ reduces to the classical Fisher information
$\cf$ and hence characterises the information geometry of the resulting probability vector via Eq.~\eqref{metric_def}
--  refer to our concise proof in Theorem~\ref{class-fisher} in Appendix~\ref{app:proofs}.

In summary, the quantum Fisher information generalises and unifies various
notions related to the geometry of quantum states. Furthermore, $\qf$ characterises
the information geometry~\cite{qfi1,morozova1991markov, petz_geometry,petz_book_chapter}
of probability distributions produced upon measurements as discussed in Sec.~\ref{qfishersec}.
Our approach is hence directly related to the well-known information-geometric natural gradient approach
from machine learning \cite{amari1997neural,goodfellow2016deep,amari2000adaptive}
and generalises it to the case of arbitrary quantum states.

\subsection{Natural gradient descent for arbitrary quantum states\label{mainresult}}

We are now equipped to propose our generalisation of the natural gradient evolution from Eq.~\eqref{naturalgradEvo}.
Our aim is, e.g., to minimise the expectation value $E(\underline{\theta}) =\tr[ \rho(\underline{\theta}) \mathcal{H} ]$
of a Hermitian observable $\mathcal{H}$ over the parameters $\underline{\theta}$
using a variational quantum circuit that depends on these parameters.
This circuit produces the quantum states $\rho(\underline{\theta})  = \Phi(\underline{\theta}) \, \rho_0$
via a mapping as discussed in Sec~\ref{circuitSec}
and might, for example, involve non-unitary transformations due to experimental imperfections
or indeed intentional non-unitary elements, such as measurements.
We now state our main result, the quantum natural gradient update rule
as a direct quantum analogue of Eq.~\ref{naturalgradEvo}.
\begin{result} \label{result1}
	The natural gradient update rule for parameters $\underline{\theta} = (\theta_1, \theta_2 \dots \theta_\nu)^T$
	of a variational quantum circuit is obtained as
	\begin{equation} \label{naturalgradEvoRESULUT}
		\underline{\theta}(t{+}1) = \underline{\theta}(t) - \kappa \, [\qf]^{-1} \underline{g}.
	\end{equation}
	The quantum Fisher information matrix $\qf$ corrects the gradient vector
	$g_k := \partial_k E(\underline{\theta})$
	to account for the co-dependent and non-uniform effect
	of the parameters on an
	\textbf{arbitrary quantum state}
	$\rho(\underline{\theta})$ (mixed or pure).
	The objective function is usually the expectation value
	$E(\underline{\theta}) :=\tr[ \rho(\underline{\theta}) \mathcal{H} ]$
	but in general it can be any smooth or at least differentiable mapping.
\end{result}

Computing the matrix $\qf$ can be involved for arbitrary quantum states
and there are numerous expressions available in the literature \cite{qfi1,qfi2}.
Here we state one expression that is valid for general rank-$r$ density matrices
$\rho(\underline{\theta}) = \sum_{n=1}^r p_n |\psi_n \rangle \langle \psi_n | $
with $p_n>0$, where both the eigenvalues $p_n:=p_n(\underline{\theta}) $ and
the eigenvectors $|\psi_n \rangle := |\psi_n (\underline{\theta}) \rangle$ may depend on the parameters
(we will consistently omit this dependence in our notation).
Matrix entries of $\qf$ are given as
\begin{align}\nonumber
	[\qf]_{kl}  &= \sum_{n=1}^r \frac{(\partial_k p_n)(\partial_l p_n)}{p_n} 
	+ 
	\sum_{n=1}^r 4 \, p_n \mathrm{Re}[\langle \partial_k \psi_n |\partial_l \psi_n \rangle]\\
	&- \sum_{n,m=1}^r \frac{8 p_n \, p_m}{p_n + p_m} \, \mathrm{Re}[
	\langle \partial_k \psi_n | \psi_m \rangle \langle  \psi_m | \partial_l \psi_n \rangle],
	\label{qfiexpreesion}
\end{align}
and recall that we use abbreviated notations $\partial_k p_n := \tfrac{\partial p_n}{\partial \theta_k}$
	for derivatives with respect to our parameters $\theta_k$.
Other general expressions or analytical solutions in special cases can be
found in, e.g., \cite{qfi1,qfi2} and we remark that our general method also
applies to infinite-dimensional quantum states as continuous-variable systems
(simplified analytical expressions for the entries
$[\qf]_{kl}$ are available for, e.g., Gaussian states in \cite{qfi1,qfi2}).
For illustration purposes we analytically solve the natural gradient evolution of a single
qubit (refer to Appendix~\ref{obj-func-sec} and to Appendix~\ref{1qb-solution}),
while we introduce general, experimentally-efficient approximations in Sec.~\ref{approxsec}.

\subsection{Natural gradient for idealised unitary circuits \label{pure-state-sec}}
As an important special case of our general approach, let us restrict ourselves now
to idealised unitary circuits as discussed in Sec.~\ref{recap}.
This special case translates to the limiting case of rank-one density matrices $r=1$
(as pure quantum states) in the general expression for $\qf$ in Eq.~\ref{qfiexpreesion}.
Our Result~\ref{result1} naturally reduces to the pure-state
variants of imaginary time evolution and natural gradient from Sec.~\ref{recap}
which used the well-known, unique metric object, the
Fubini-Study metric tensor~\cite{study1905kurzeste,fubini1904sulle,wilczek1989geometric,hackl2020geometry}.
\begin{remark} \label{remark1}
	For pure quantum states $\rho(\underline{\theta}) =| \psi \rangle \langle \psi|$
	(i.e., circuits composed of idealised unitary gates),
	the gradient of an expectation value simplifies to $\partial_k\tr[ \rho(\underline{\theta}) \mathcal{H} ]
	= \partial_k \langle \psi| \mathcal{H}| \psi \rangle$.
	As discussed in Sec.~\ref{geometry_sec}, $\qf$ reduces to
	the Fubini-Study metric tensor $\matra$ from Eq.~\eqref{imagtimeUpdate} as
	\begin{equation*}
		[\qf]_{kl}  = 4 \mathrm{Re}[\langle \partial_k \psi | \partial_l \psi \rangle 
		- \langle \partial_k \psi | \psi \rangle \langle  \psi | \partial_l \psi \rangle]
		= 4 [\matra]_{kl}.
	\end{equation*}
	It follows that the natural gradient update rule in Result~\ref{result1} with a step size
	$\kappa = 4 \, \Delta t$ is identical to a simulated
	imaginary time evolution of the pure state vector $| \psi \rangle$ as in \cite{samimagtime,xiaotheory}.
\end{remark}
We remark that previous techniques from~\cite{samimagtime,xiaotheory} are applicable for experimentally
estimating entries of the metric tensor $[\qf]_{kl} =  4 [\matra]_{kl}$ for the above unitary
pure-state evolutions.
Furthermore, when the state vector is isomorphic to a classical probability distribution
then one has $\cf = \qf$ and our Result~\ref{result1} further reduces
to the classical-Fisher-information
based approach known from machine learning \cite{amari1997neural,goodfellow2016deep,amari2000adaptive}.
Refer to our concise proof in Theorem~\ref{class-fisher} (\spl).

\subsection{Possible generalisations and improvements \label{obj-func-sec}}

Our approach in principle allows for the following generalisations and improvements relative 
to imaginary time evolution \cite{samimagtime,xiaotheory}
and the pure-state variant of natural gradient \cite{quantumnatgrad}
which are defined for objective functions of the form
$E(\underline{\theta}) := \tr[| \psi(\underline{\theta}) \rangle \langle \psi(\underline{\theta}) | \mathcal{H}]$.

First, even when the objective function is generated by an observable as $E(\underline{\theta}) = \tr[\rho(\theta) \mathcal{H}]$,
our definition in Sec.~\ref{circuitSec}
allows for general non-unitary elements as CPTP maps
which in principle enable us the manipulation of exponentially more degrees
of freedom:
We prove in Theorem~\ref{Fourier_series_theorem} (refer to the \spl)
that in general the increased dimensionality gives rise to significantly more expressive objective functions.
For example, a unitary circuit consisting of $\nu$ single- or two-qubit gates
can be expanded into $\mathcal{O}(2^{4\nu})$ Fourier components while
a similar circuit of single- or two-qubit CPTP maps
results in $\mathcal{O}(2^{8\nu})$ terms. 
While we show in Appendix~\ref{effective-cptp-map} that general local CPTP maps can be implemented efficiently,
in the rest of this work we focus on the special but pivotal case when non-unitary of the gates
is due to hardware imperfections and derive an efficient approximation of the QFI matrix.

Second, the expected value $E(\underline{\theta}) := \tr[| \psi(\underline{\theta}) \rangle \langle \psi(\underline{\theta}) | \mathcal{H}]$ is a mapping that is linear in the quantum state. 
It was recently shown that in case of Pauli gates (most typical circuit construction) these objective functions
$E(\underline{\theta})$ can have relatively simple structure, i.e., they can be be specified as simple trigonometric series \cite{koczor2022quantum}.
In contrast, our Result~\ref{result1} is well-defined for more general
objective functions and its convergence is guaranteed for the general class
of mappings that give rise to Lipschitz-continuous gradients --
even in the presence of shot noise \cite{sweke2019stochastic}.
For example, such continuously differentiable, non-linear functions include, among others, polynomials 
$E(\underline{\theta}) = \mathrm{poly}(x_1,x_2, \dots x_n)$ 
that depend on a set of expectation values $x_k = \tr[\rho(\theta) \mathcal{H}_k]$,
or analytic functions of $x_k$.
One could also consider, e.g., the cross entropy, the log-likelihood functions or beyond.
In particular, the metrological performance
of a quantum state was recently used as an objective function \cite{koczor2019variational},
and our natural gradient approach could be an invaluable tool for solving such non-trivial optimisation problems.

Finally,
we analytically solve the parameter evolution of a single qubit in Appendix~\ref{1qb-solution}
to illustrate that the metric information becomes particularly important in the case of non-unitary evolutions;
here we only recollect the most important points.
First, the metric tensor $\qf$ becomes singular when approaching pure states
(see division by vanishing probabilities in Eq.~\eqref{qfiexpreesion}) and small variations in the parameters can
result in increasingly large `jumps' in state space. 
As such, the metric tensor in our approach `regulates' those ill-behaved parameter evolutions.
Second, our example in Appendix~\ref{1qb-solution} nicely demonstrates the previous general point
and illustrates that the metric information in consequence improves exponentially on
the convergence speed of simple gradient descent.
Third, the simple gradient evolution in general highly depends on the parametrisation
and gets trapped in local minima. In contrast, the natural gradient evolution
does not depend on the parametrisation and avoids local minima,
see also \cite{samimagtime, wierichs2020avoiding}.

\section{Efficient approximation for experimental implementation} \label{approxsec}
\subsection{Experimentally estimating the QFI matrix}
Our numerical simulations in Sec.~\ref{numsim} used the exact
quantum Fisher information matrix $\qf$ whose numerical computation 
in general is involved.
We now consider efficient schemes for approximating $\qf$ in case when
the non-unitarity of the ansatz circuit is due to small, but non-negligible
imperfections of individual quantum gates, as in case of NISQ hardware.
Let us also note that the $\qf$ is typically ill-conditioned and
approximating its inverse in the update rule in Result~\ref{result1}
(using, e.g., regularisation techniques) unavoidably introduces an error
--
especially under experimental shot noise~\cite{van2020measurement}.
For practical purposes one is thus content with an approximation of the quantum Fisher information.

Let us first restrict the process in Sec.~\ref{circuitSec}
to model the effect of experimental imperfections
that are typical to near-term quantum hardware.
For this reason, we adopt the analysis of
refs.~\cite{koczor2021dominant,koczor2020exponential,huggins2020virtual}
of typical error models encountered in noisy, near-term quantum devices.
In particular, we write mixed quantum states in terms of their spectral decomposition as
\begin{equation}
	\rho := \lambda |\psi \rangle \langle \psi | +  \sum_{m=2}^d \lambda_m |\psi_m \rangle \langle \psi_m |.
\end{equation}
It has been shown that quantum hardware are expected to produce mixed quantum states whose
dominant eigenvector $|\psi \rangle $ approximates the ideal computational quantum
state due to the high entropy of the error eigenvalues (probabilities) $\lambda_m$
-- and this entropy increases when we increase the system size, refer to~\cite{koczor2021dominant,koczor2020exponential,huggins2020virtual} for more details.
Furthermore, in the limit of maximal entropy as
$\lambda_2 = \lambda_3 \dots = \lambda_d = (1-\lambda)/d$, the
error eigenvalues decrease exponentially when we increase the system size $d=2^N$ and
the dominant eigenvector is identical to the ideal computational state.
We will refer to this limit in the following as global depolarising noise.
The ESD/VD techniques 
build on the above arguments and use multiple copies of a mixed quantum state $\rho$
to estimate expectation values in the dominant eigenvector $|\psi_1 \rangle \equiv |\psi \rangle$ of the noisy quantum state
with an exponentially decreasing error (in the number of copies).

An important ingredient to our approach is that we can efficiently estimate the metric tensor  $\tr[(\partial_k \rho)(\partial_l \rho)]$
as Hilbert-Schmidt scalar products~\cite{ReedSimon1} between partial derivatives of the density operator.
We note that corresponding SWAP-test circuits were proposed for variational simulations of imaginary time evolution of
mixed states in Eq.~\ref{matmdef} via ref.~\cite{xiaotheory} and we discuss in Appendix~\ref{densityimag}
that this metric tensor is generally different than the QFI.

As such, we prepare two copies of the experimental quantum
state $\rho$ to estimate the Hilbert-Schmidt metric tensor $\tr[(\partial_k \rho)(\partial_l \rho)]$,
either by applying controlled-gate generators as in~\cite{xiaotheory}
or via parameter shifts as in \cite{koczor2022quantum}.
The resulting approach is therefore  a special case of the ESD/VD technique which 
we exploit to show that we can obtain an excellent approximation of the
QFI of the dominant eigenvector $|\psi \rangle $ of
the noisy quantum state in a comprehensive range of NISQ scenarios.

\begin{result} \label{result3}
The Hilbert-Schmidt metric tensor of an arbitrary mixed quantum state $\rho$ as $\tr[(\partial_k \rho)(\partial_l \rho)]$
can be estimated experimentally using the same resources as of
the ESD/VD technique~\cite{koczor2020exponential,huggins2020virtual}.
This metric tensor approximates the quantum Fisher information
of the dominant eigenvector $|\psi \rangle$ of the quantum state as
\begin{equation*}
	\tr[(\partial_k \rho)(\partial_l \rho)]
	=
	\frac{\lambda^2}{2} [\qf(|\psi\rangle\langle\psi|)]_{kl} + E,
\end{equation*}
up to a bounded error $|E|	\leq   a +  b \lambda_2$. Here the term
$a:= \max_{k} [\sum_{m=1}^d |\partial_k \lambda_m|^2]$ expresses how rapidly the eigenvalues of $\rho$ can change under the parametrisation
and it is expected that $a\ll 1$ in typical NISQ scenarios
while $b$ is constant bounded. For example, if the
parametrisation corresponds to Pauli gates (most common universal construction)
then $b\leq 1$, which we discuss in  Definition~\ref{definition:bterm}.
Note that the multiplier $\lambda^2/2 \leq 1/2$
can be omitted as it only multiplies the step size $\kappa$
by a constant bounded factor, and thus only trivially affects an optimisation.
\end{result}

\begin{figure}[tb]
	\begin{centering}
		\includegraphics[width=0.45\textwidth]{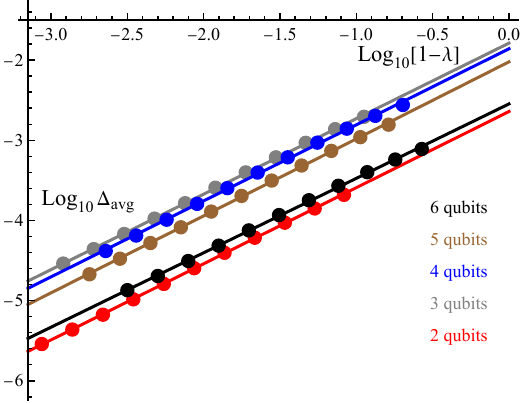}
		\caption{
			Average approximation error $\Delta_{\rm{avg}}$ of the quantum
			Fisher information of a noisy quantum state $\rho$  under \emph{local} depolarising noise
			(solid circles) was computed assuming the experimental system in Sec.~\ref{numsim}
			at different local gate noise $10^{-4} \leq p_{err} \leq 10^{-2}$.
			The average errors depend linearly on the dominant eigenvalue (solid lines) as
			$\mathcal{O}(1{-}\f)$ while the absolute error decreases rapidly (exponentially for \emph{global} depolarising noise)
			when we increase the numbers of qubits (parallel lines in the log-log plot).
			These numerical observations reinforce that our general Result~\ref{result3}
			applies well under experimental conditions even in the specific instance when we
			assume global depolarising noise -- in which limit we expect an exponential decay
			$\Delta_{\rm{avg}}=\mathcal{O}[(1{-}\f)/d]$ via the dimensionality $d=2^N$
			as explained in the main text.
			\label{approximations}
		}
	\end{centering}
\end{figure}

The efficacy of the ESD/VD technique was shown in refs.~\cite{koczor2020exponential,koczor2021dominant,huggins2020virtual}
to crucially depend on the dominant error eigenvalue $\lambda_2$ and based on their analytical and numerical analyses
it can be expected that $\lambda_2$ decreases when we increase the system size in typical NISQ scenarios.
As such, in the above result we establish that the approximation error $|E| = a  + \mathcal{O}(\lambda_2)$
generally scales with this error eigenvalue up to a small additive error $a \ll 1$.
In fact, we show in Definition~\ref{definition:bterm} that in the case of unitary parametrisations
of arbitrary mixed sates as $U(\underline{\theta}) \rho U(\underline{\theta})^\dagger$
we are guaranteed that $a=0$.
Furthermore, we now show that in the limiting case of global depolarising noise
$a=0$ and we can estimate the QFI of both the mixed quantum state $\rho$ and the ideal computational state
up to an error that decreases exponentially with the number of qubits.

\begin{result}\label{resultTHREE}
	In the limit of global depolarising noise we obtain the simplified error bounds
	from Result~\ref{result3} as
	\begin{equation*}
		\tr[(\partial_k \rho)(\partial_l \rho)] =
		\frac{\lambda^2}{2} [\qf(|\psi\rangle\langle\psi|)]_{kl}	+ E_{depol},
	\end{equation*}
	with an approximation error  $|E_{depol}|	\leq    b (1-\lambda)/d$
	that scales inversely with the
	dimensionality $d=2^N$.
	Furthermore, we can also approximate the QFI
	of a mixed quantum state (and not just its dominant eigenvector) as
	\begin{equation*}
		\tr[(\partial_k \rho)(\partial_l \rho)] =	\frac{\lambda}{2} [\qf(\rho)]_{kl} + E'_{depol}.
	\end{equation*}
	Here the approximation error $E'_{depol} \in \mathcal{O}[(1-\lambda)/d]$ similarly scales inversely with the
	dimensionality.
\end{result}

Global depolarising noise
	has been rigorously proven to be a very good approximation in random quantum circuits~\cite{dalzell2021random}
	and it is commonly assumed that experimental noise models can be approximated well by global depolarisation
as in \cite{PhysRevE.104.035309,arute2019quantum}. This approximation can be expected to enhance when we
increase the system size via refs~\cite{koczor2021dominant, koczor2020exponential, huggins2020virtual}
which we numerically validate assuming a typical experimental noise model:
We plot numerically computed average approximation errors of the QFI in Fig.~\ref{approximations}
for different values of dominant eigenvalues $1{-}\f$ as obtained by increasing the strength of
local depolarising noise in an experimental quantum system.
These errors show a linear trend as a function of $1{-}\f$ (for a fixed number of qubits -- horizontal direction).
Furthermore, our bounds assuming global depolarising noise apply particularly well
as the approximation error from Result~\ref{result3} decreases rapidly as we increase the system size.
In particular, errors in Fig.~\ref{approximations} decrease for a fixed
dominant eigenvalue $\f$ when we increase the number of qubits beyond $N=2$ (decreasing parallel lines).
Most importantly, the absolute values of errors $\Delta_{\rm{avg}} < 10^{-3}$ are expected to
be negligible in practically relevant scenarios, e.g., when $\f > 0.1$, especially when compared to shot noise or
other sources of practical errors (e.g., regularisation).

\begin{figure*}[tb]
	\begin{centering}
		\includegraphics[width=0.85\textwidth]{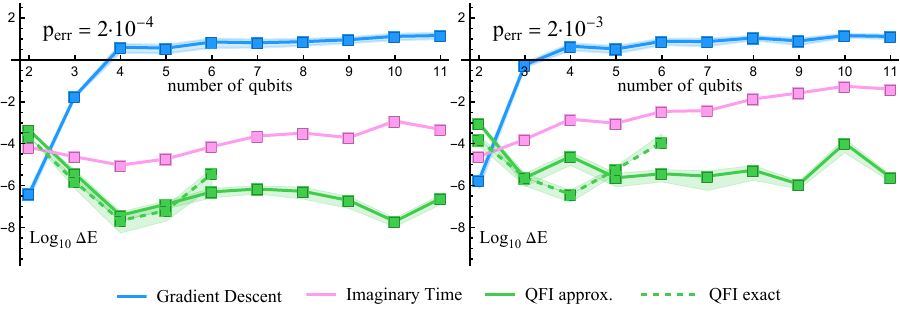}
		\caption{
			Comparing the performance of various optimisation techniques as a function of the qubit number $N$
			on the problem of finding the minimal energy $E_{\mathrm{opt}}$ for the Hamiltonian
			in Eq.~\eqref{spin_ring} using noisy variational circuits. Gradient descent (blue), QFI-based
			natural gradient (green)
			and imaginary time evolution of mixed states (pink) were started from the same randomly chosen initial positions near the optimum.
			Logarithmic distance from the optimal energy $E_{\mathrm{opt}}$
			after an evolution of 60 steps assuming different
			severity $p_{err}$ of errors.
			Performance of the optimisation is very similar with our efficient approximate (green solid) QFI matrix
			as with the exact one (green dashed).
			\label{approxsimul}
		}
	\end{centering}
\end{figure*}

We note that in Fig.~\ref{approximations} we are limited to simulations of only a few qubits as $N\leq 6$
since computing the QFI exactly is inefficient. Nevertheless, in the Appendix in Fig.~\ref{fig:aterms} we
provide comprehensive numerical evidence that the additive error in our upper bounds is small as $a \ll 1$.
This further supports that our bounds assuming global depolarising noise apply particularly well
when we increase the system size as the error eigenvalues $\lambda_m$ approach close to maximum entropy.
Furthermore, in Fig.~\ref{introfig}(b) (white) we
illustrate how impressively well Result~\ref{result3} approximates the exact evolution.

Let us finally note that in general $\tr[(\partial_k \rho)(\partial_l \rho)]$
is a Riemannian metric tensor that is obtained in the Hilbert-Schmidt metric
\cite{Sommers_2003}. As such, the above result establishes that the geometry of specific quantum
states produced by near-term quantum hardware, as characterised by the quantum Fisher information,
can be well approximated by the Hilbert-Schmidt metric -- even though the two geometries can generally
be very different \cite{Sommers_2003}.
We remark that our general approximation in Result~\ref{result3} applies to infinite-dimensional
density matrices too as continuous-variable systems.
We further remark that the update rule via Result~\ref{result1} using our approximation of $[\qf]_{kl} $
is fundamentally different from imaginary time evolution (of mixed states from Sec.~\ref{densityimag} in the \spl).
In particular, our natural gradient approach uses the above matrix as an approximation of the quantum Fisher
information which is then used to correct the gradient vector $g_k =  \partial_k E(\underline{\theta})$
to account for the non-uniform and co-dependent effect of the parameters on quantum states.
Fig.~\ref{introfig} (pink) illustrates
how their trajectories are different and imaginary time evolution gets stuck
at a point away from the optimum where its `gradient' vector $\underline{Y}$ vanishes.

\subsection{Numerical validation \label{numsim}}

	We consider the spin-ring Hamiltonian
	in Eq.~\eqref{spin_ring} and compare our approach to other optimisation techniques using a
	noisy ansatz circuit for an increasing number of qubits.
	In Fig.~\ref{approxsimul} we show the distance $\Delta E$ from the optimal energy $E_{\mathrm{opt}}$
	after we evolve for a fixed number $60$ of iterations starting at random positions around the optimum
	--
	we thereby validate our approach based on its performance in the final
	stages of an optimisation where differences in convergence speed are most pronounced.
	Solid lines show the average distance achieved over 25 runs
	while shading represents the standard deviation.
	Fig.~\ref{approxsimul} (green and pink) show that
	imaginary time 
	evolution (of mixed states as in Appendix~\ref{densityimag})
	and our quantum-Fisher-information based method get significantly closer to the optimum
	than gradient descent -- albeit imaginary time evolution 
	converges to a point slightly away from the optimum as
	also illustrated in Fig.~\ref{introfig}.

Fig.~\ref{approxsimul} (green solid) illustrates that the error introduced by our efficient approximation of 
the QFI matrix does not significantly affect the performance of the optimisation using the exact QFI matrix
in Fig.~\ref{approxsimul} (green dashed).
We also note that the Hilbert-Schmidt scalar products in Result~\ref{result3} are more
efficient to compute in a quantum simulator
than numerically evaluating the exact QFI in Eq.~\eqref{qfiexpreesion};
the software package Quantum
Exact Simulation Toolkit (QuEST) and in QuESTlink\,\cite{QuESTlink}
have built-in functionality for computing such scalar products which
	enabled us to simulate a significantly larger number of
	qubits in Fig.~\ref{approxsimul} (green solid) than in (green dashed).

\subsection{Measurement overhead analysis}
In our numerical simulations we focused on the fundamental limitations posed by hardware noise.
	We now discuss that shot noise due to finite sampling affects the performance of our approach, primarily
	as the signal to be estimated, i.e., as characterised by the dominant eigenvalue $\lambda$, decreases
as we increase the number of imperfect gates in a quantum circuit~\cite{koczor2021dominant}.
For instance, if a circuit
acts on $N$ qubits with $N_g$ gates, each of which is characterised
by an independent error probability $p_{err}$, then the resulting eigenvalue can be approximated
as $\f \approx (1-p_{err})^{N_g} \approx e^{- \xi} $, where $\xi := p_{err} N_g$ is the circuit error rate.
This exponential decay of $\f$ results in
a decay of the term
$\tr[(\partial_k \rho)(\partial_l \rho)] = \mathcal{O}(\f^2) = \mathcal{O}( e^{- 2\xi} )$
in Result~\ref{result3}
 and affects our approach in the following ways.

a) The decreasing term $\tr[(\partial_k \rho)(\partial_l \rho)] = \mathcal{O}(\f^2)$
is estimated by measuring an ancilla qubit.
This requires a measurement overhead $N_s = \mathcal{O}( \f^{-4})$ 
to sufficiently reduce statistical fluctuations caused by shot noise, as in the case of the ESD/VD techniques.
This is not a significant increase in the number of measurements
for moderate values of infidelities, i.e., less than a factor of $16$ for $\f>0.5$.

b) Keeping the measurement overhead constant poses a limitation on the maximum gate count via 
$N_g^{\mathrm{max}}= \ln(N_s)/[-2\ln(1-p_{err})]$.
We illustrate this on a simple example: take a linear number of gates in the circuit $N_g = a N$
and assume that two-qubit gates are the dominant source of errors with an effective $a\approx3$, e.g., as in
Fig.~\ref{circ}. For a constant measurement overhead $N_s = 16$,
the maximum number of qubits would be $N^{\mathrm{max}} \approx 230$ when  $p_{err} =10^{-3}$
and $N^{\mathrm{max}} \approx 2310$ when $p_{err} =10^{-4}$, where the latter corresponds to state-of-the-art
two-qubit gate errors.

c) As the number of qubits increases $N \gg 1$, the approximation error $\mathcal{O}(2^{-N})$ 
assuming global depolarisation
decays faster in the number of qubits than the quantity to be measured
$\f^2 = [(1-p_{err})^{2 N_g/N}]^{N}$ when circuits consist of a  linear number of gates
$N_g^{\mathrm{max}} = a N$ with depth $a < -\ln(2)/[2\ln(1-p_{err})]$
and for small $p_{err}$ we approximately have $a \lessapprox \ln(2)/(2p_{err})$.
This translates to depth
$a = 346$  when $p_{err} = 10^{-3}$ and depth $ a =3465$ when $p_{err} =10^{-4}$.
On the other hand, for super-linearly scaling gatecounts the relative error
decays until reaching a maximal number qubits $N^{\mathrm{max}}$ -- even though it
diverges asymptotically. To illustrate this, we take an example of a logarithmic-depth
circuit $N_g = a N \log N$ in  which case the relative error decreases until
an optimal number of qubits $N^{\mathrm{max}} =\exp\{-\ln(2)/[2 a \ln(1 - p)]-1\}$ is reached.
This results in $N^{\mathrm{max}} = 4.1 \times 10^{14}$ when $a = 10$ and $p_{err} = 10^{-3}$
and results in $N^{\mathrm{max}} = 4.6 \times 10^{29}$ when $a = 50$ and $p_{err} = 10^{-4}$.
This example illustrates that
the relative error is expected to decrease in the regime relevant in the context of NISQ.

In summary, our approach in Result~\ref{result3} is robust to errors
and its scalability is primarily limited by the circuit error rate, i.e., the expected number
of errors $\xi$ in a circuit execution must not significantly exceed $1$.
We further note that similar limitations hold
for estimating the gradient vector and the cost function, e.g., the measurement
overhead scales as $N_s = \mathcal{O}(\f^{-4})$ when using the ESD/VD techniques with two copies
	of the experimental quantum state~\cite{koczor2020exponential,huggins2020virtual,koczor2021dominant}.

Let us finally compare Result~\ref{result3} to the experimental protocol that
simulates pure-state imaginary time evolution (as discussed in Sec.~\ref{recap}) \, \cite{samimagtime,xiaotheory}.
This approach estimates the entries $[\matra]_{kl}$ of the pure-state metric tensor
	using Hadamard-test circuits.
Although these circuits require only a single copy of the quantum state, the
	circuit depth is nearly double that of the circuit depth of our approach.
	For this reason we expect that similar limitations on the gate count and measurement
	overhead apply here. 
	However, these Hadamard-test circuits yield a non-trivial, non-negligible error when run on imperfect quantum 
	hardware as their approximation error $\mathcal{O}(1{-}\f)$
	may be more substantial than the error term in Result~\ref{result3} which is suppressed for an increasing system size (exponentially
	in the limit of global depolarising noise).
	Indeed, comparing yellow and white evolutions in Fig.~\ref{introfig} illustrates that
	Result~\ref{result3} provides a significantly more error-robust
	approximation of the optimal evolution in practice.

\section{Discussion and Conclusion}

In this work we have extended the quantum natural gradient approach from the well-studied
unitary (and therefore noise-free) quantum circuits to the most general scenario of optimising arbitrary quantum states as density matrices.
We have shown that the quantum Fisher information, a quantity much-studied in the context of quantum metrology, can be used to
correct the gradient vector in a variational quantum algorithm to account
for the non-uniform effect of the parameters on the underlying mixed quantum states.

As the main practical result,
we devised an efficient, noise-aware experimental protocol for estimating
the quantum Fisher information matrix
in case when non-unitarity of ans{\"a}tze is due to small experimental imperfections of
the quantum gates.
Our approach is closely related to the ESD/VD error suppression technique~\cite{koczor2020exponential,huggins2020virtual}
and also fits well with a multicore architecture~\cite{jnane2022multicore}. Building on prior theoretical results on the dominant eigenvector~\cite{koczor2021dominant} of quantum states we rigorously proved
that we can approximate the QFI in typical near-term quantum circuits via the Hilbert-Schmidt metric tensor.
This result also establishes the fundamental observation that typical noisy quantum states produced by near-term
quantum devices have an (approximately) identical geometry in either the Hilbert-Schmidt metric or in terms
of the QFI.

We assessed the practical usefulness of our approach and showed that it performs impressively
well in typical practical scenarios. A significant advantage of our technique is that it can be used
in tandem with exponentially powerful error mitigation schemes
and as we proved it inherits similarly strong theoretical guarantees~\cite{koczor2020exponential,huggins2020virtual}.
In particular, our error term in Result~\ref{result3} only depends on the eigenvalue-distribution of
the density matrix and is oblivious to the particular error model of the quantum device -- and requires no explicit knowledge thereof.
The main limitation is indeed a mismatch in the  dominant eigenvector
and our numerical experiments confirmed prior expectations~\cite{koczor2020exponential,huggins2020virtual}
that this coherent mismatch gets suppressed very well by a variational minimisation,
e.g., errors below $\Delta E \leq 10^{-4}$ in Fig.~\ref{fig:vqe_plots}.

When compared to previous studies, our approach has the advantage that it explicitly takes into account imperfections
of the variational quantum circuit. 
It is therefore appropriate for seeking the optimum when the quantum circuits to be employed are imperfect, and this has
been the focus of our numerical simulations and our study of approximation methods. However we emphasise that the applicability
of the method is not restricted to noisy unitary circuits, but can be applied to the far-reaching scenario when a circuit contains
intentional non-unitary transformations, such as measurements or variable-time decoherence. The implications of this flexibility
form an interesting topic for future work.

\section*{Acknowledgments}
S.~C.~B. acknowledges financial support from the EPSRC Hub grants under the grant agreement numbers as
EP/M013243/1 and EP/T001062/1, and from the IARPA funded LogiQ project.
B.\,K. and S.\,C.\,B. acknowledge funding received from EU H2020-FETFLAG-03-2018 under the grant
agreement No 820495 (AQTION).
B.K. thanks the University of Oxford for
a Glasstone Research Fellowship and Lady Margaret Hall, Oxford for a Research Fellowship.
The numerical modelling involved in this study made use of the Quantum Exact Simluation Toolkit (QuEST), and the recent development QuESTlink\,\cite{QuESTlink} which permits the user to use Mathematica as the integrated front end. The authors are grateful to those who have contributed to both these valuable tools. 
The views and conclusions contained herein are those of
the authors and should not be interpreted as necessarily representing the official policies or endorsements, either expressed or implied, of the ODNI, IARPA, or the
U.S. Government. The U.S. Government is authorized to
reproduce and distribute reprints for Governmental purposes notwithstanding any copyright annotation thereon.
Any opinions, findings, and conclusions or recommendations expressed in this material are those of the author(s) and do not necessarily reflect the view of the
U.S. Army Research Office.


%

\appendix
\onecolumngrid

\section{Imaginary time evolution of mixed quantum states \label{densityimag}}
Let us consider now the imaginary time evolution of
an initial \emph{mixed} quantum state $\rho$ as 
\begin{equation*}
	\rho(t) = e^{- \mathcal{H} t } \rho e^{- \mathcal{H} t } / \tr[ e^{-2 \mathcal{H} t } \rho].
\end{equation*}
This evolution increases or decreases mixedness of the density operator and,
as a consequence, it does not
reduce to the pure-sate imaginary time evolution (from Eq.~\eqref{imagtimeUpdate} in the main text)
in the limiting case of pure states.
We will consistently refer to this update rule as imaginary time evolution
of mixed states or density matrices (as opposed to state vectors $| \psi \rangle$
discussed in Sec.~\ref{recap} in the main text).

It has been shown in \cite{xiaotheory} that the closest \emph{unitary evolution} can be
simulated efficiently using variational quantum circuits.
We assume that these circuits produce quantum states via a mapping
$\rho:= \rho(\underline{\theta})$ as discussed in Sec~\ref{circuitSec}.
The corresponding
parameter-update rule is analogous to Eq.~\eqref{imagtimeUpdate}
and results in \cite{xiaotheory}
\begin{equation}\label{densityimagtimeUpdate}
	\underline{\theta}(t{+}1) = \underline{\theta}(t) + \Delta t \, \matrm^{-1} \underline{Y},
\end{equation}
but here the vector $\underline{Y}$ appears instead of the energy gradient
$g_k =  2 \mathrm{Re} \{ \tr[ (\partial_k \rho) \mathcal{H} ] \}$
and its entries have the explicit form $\underline{Y}_k = - \mathrm{Re} \{ \tr[ (\partial_k \rho) \mathcal{H}\rho ] \}$.
The matrix $\matrm$ contains Hilbert-Schmidt scalar products
between differentials of the mixed state
\begin{equation}\label{matmdef}
	[\matrm]_{kl} = \tfrac{1}{2}\tr[ (\partial_k \rho) (\partial_l \rho)].
\end{equation}

Although this approach results in an improved performance when compared
to simple gradient descent -- see Fig.~\ref{introfig} (pink) --
it is vulnerable to becoming stuck in a point away from the optimum.
This is because the exact gradient of imaginary time evolution is non-zero but
points to a purely non-unitary direction.
Furthermore, this approach does not reduce to the previously discussed
pure-sate imaginary time evolution from Eq.~\ref{imagtimeUpdate} in the
limiting case of pure states.

In particular, the matrix $\matrm$  reduces to
the quantum Fisher information and to the Fubini-Study metric tensor
as $ \qf = 4 \matrm = 4\matra$ in the limiting case of our Result~\ref{result3}
for pure states via $\epsilon\rightarrow0$. However, the vector entries
$\underline{Y}_k$ do not reduce to the pure-state gradient vector $g_k = \partial_k\tr[ \rho(\underline{\theta}) \mathcal{H} ]
= \partial_k \langle \psi| \mathcal{H}| \psi \rangle$. We show this via a direct calculation
\begin{align*}
	\underline{Y}_k &= - \mathrm{Re}\,  \tr[ (\partial_k \rho) \mathcal{H}\rho ] 
	= - \mathrm{Re} \,   \tr[ ( |\partial_k \psi \rangle \langle \psi | + | \psi \rangle \langle \partial_k \psi | ) \mathcal{H}  | \psi \rangle \langle \psi |  ] \\
	&=
	- \mathrm{Re} \,  \tr[  |\partial_k \psi \rangle \langle \psi |  \mathcal{H}  | \psi \rangle \langle \psi |  ]
	- \mathrm{Re}  \, \tr[  | \psi \rangle \langle \partial_k \psi | \mathcal{H}  | \psi \rangle \langle \psi |  ] \\
	&= 
	- E \, \mathrm{Re} \,  \tr[  |\partial_k \psi \rangle  \langle \psi |  ]
	- \mathrm{Re}  \, \langle \partial_k \psi | \mathcal{H}  | \psi \rangle.
\end{align*}
Here the second term is proportional to the expected gradient as
$- \mathrm{Re}  \, \langle \partial_k \psi | \mathcal{H}  | \psi \rangle = - g_k/2$
while the first term can be simplified as $- E \, \mathrm{Re}  \langle \partial_k \psi | \psi \rangle  $, which only
vanishes when there is no global phase evolution under a variation of parameters.
We conclude by recollecting the above  	terms as
\begin{equation*}
	\underline{Y}_k = - g_k/2 - E \, \mathrm{Re}  \langle \partial_k \psi | \psi \rangle, \quad \quad \text{when}
	\quad \quad \rho = | \psi \rangle \langle \psi |.
\end{equation*}

In summary, the above mixed-state imaginary time evolution has three main drawbacks.
First, it uses the metric tensor $\matrm$  which is obtained in the Hilbert-Schmidt metric.
As such, the Hilbert-Schmidt metric results in a different
geometry when compared to the $\qf$ considered in our work:
While $\qf$ is both Riemannian and monotone,
$\matrm$ is Riemannian but not monotone under CPTP maps
\cite{Sommers_2003}. 
Second, the mixed-state imaginary time evolution does not completely
reduce to the corresponding pure-state variants.
Third, it does not necessarily converge to the true optimum.

\begin{figure*}[tb]
	\begin{centering}
		\includegraphics[width=0.7\textwidth]{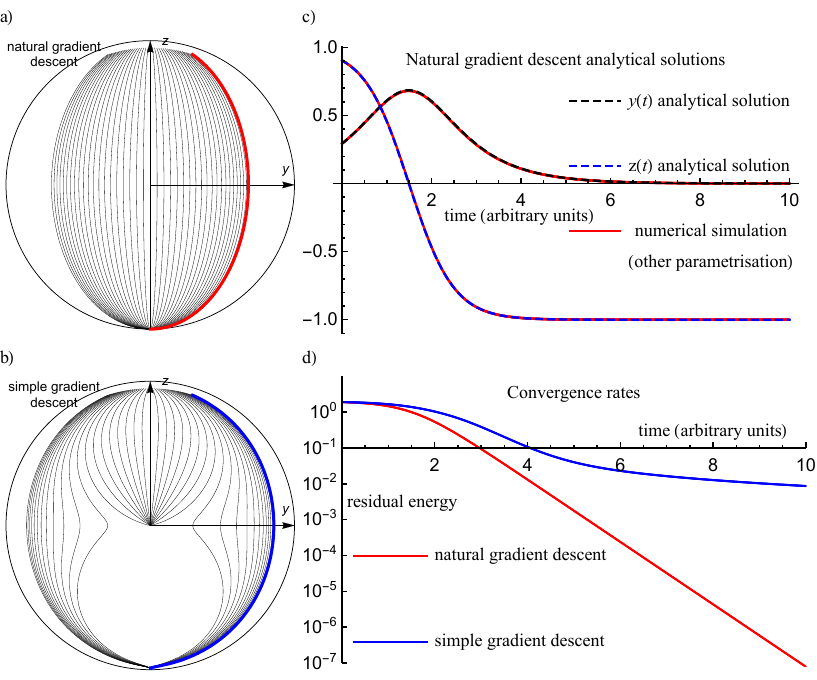}
		\caption{
			a) Analytical solution of the natural gradient evolution
			from Eq.~\eqref{descent-solution}
			of a single qubit shown via its Bloch representation.
			The solution is independent of the parametrisation:
			red curve shows numerical simulations in a different, polar parametrisation.
			b) comparison of the analytical solution and the numerical simulation
			form a) which use different parametrisations.
			c) Numerical simulation of simple gradient descent in a polar parametrisation.
			The evolution strongly depends on the parametrisation and gets trapped 
			in local minima.
			d) The natural gradient evolution converges exponentially faster to the
			desired solution than simple gradient descent.
		}
		\label{analyitc-fig}
	\end{centering}
\end{figure*}

\section{Analytically solvable case of single qubits}\label{1qb-solution}

We now consider the explicit example of an arbitrary single-qubit
state $\rho$ and analytically solve its natural gradient evolution.
We assume that the aim is to minimise the expectation value
$\tr[\rho \mathcal{H}]$ with respect to an arbitrary single-qubit observable $\mathcal{H}$.
We parametrise our initial state $\rho_0$ in terms of the usual Bloch representation
as a point in the Bloch sphere $\overrightarrow{r_0} := (x_0,y_0,z_0)$ with
$\lvert \overrightarrow{r_0} \rvert \leq 1$.

The identity contribution in the observable can be discarded (it does not effect
the minimisation) and we only consider the traceless part of $\mathcal{H}$. In this case the
observable can also be identified uniquely with its Bloch-vector representation $(x,y,z)$ and
we can apply a similarity transformation to our coordinate system such that the observable
is along the $z$ axis via $\overrightarrow{r}_\mathcal{H}:= (0,0,1)$ and the initial state
is in the plane $(0,y_0,z_0)$. Hence in the following we can drop the dependence on the coordinate $x$.

The cost function takes up the form $E(y,z) = \tr[\rho(y,z) \mathcal{H}] = z$ and the
gradient vector can be computed straightforwardly as $\underline{g}(y,z) = (\partial_y E(y,z), \partial_z E(y,z)) = (0,1)$.
We now use the explicit formula of the quantum Fisher information of a single qubit
Bloch vector from \cite{qfi1} as
\begin{equation*}
	[\qf]_{yz} = 
	(\partial_y \overrightarrow{r}) \cdot  (\partial_z \overrightarrow{r})  +
	\frac{
		(\overrightarrow{r} \cdot \partial_y \overrightarrow{r}) (\overrightarrow{r} \cdot \partial_z \overrightarrow{r})
	}{1- \lvert \overrightarrow{r} \rvert^2}.
\end{equation*}
We can analytically compute the explicit form of the metric tensor using the above formula
and its inverse follows as
\begin{equation*}
	[\qf]^{-1} = 
	\left(
	\begin{array}{cc}
		\frac{z^2-1}{y^2+z^2-1} & -\frac{y z}{y^2+z^2-1} \\
		-\frac{y z}{y^2+z^2-1} & \frac{y^2-1}{y^2+z^2-1} \\
	\end{array}
	\right)^{-1} 
	=
	\left(
	\begin{array}{cc}
		1-y^2 & -y z \\
		-y z & 1-z^2 \\
	\end{array}
	\right).
\end{equation*}
The above matrix is clearly singular in the limiting case of
pure states when $y^2+z^2 = 1$.
The resulting natural gradient vector is 
$\underline{g}_n = (-y z, 1 - z^2)$.

The (continuous) natural gradient evolution is generated by the set of
differential equations from Result~\ref{result1} as
\begin{align*}
	\partial_t y(t) = y(t) z(t), \\
	\partial_t z(t) =  z(t)^2 - 1. \nonumber
\end{align*}
We solve the above differential equations analytically
in terms of the initial state $\overrightarrow{r_0} := (y_0,z_0)$
\begin{align} \label{descent-solution}
	y(t) = \frac{y_0}{\cosh (t)-z_0 \sinh (t)}\\
	z(t) = -\tanh[t - \mathrm{arctanh}(z0)]
\end{align}

We plot this analytical solution on the Bloch sphere in Fig.~\ref{analyitc-fig}(a)
for a set of different initial conditions $\overrightarrow{r_0} = (y_0,z_0)$ and
compare it to the simple gradient evolution on Fig.~\ref{analyitc-fig}(b).
For the case of simple gradient descent, we have numerically simulated the evolution
in terms of the polar parametrisation $\overrightarrow{r}=(r(c),\theta) $,
where the radius is expressed as $r(c):= 1-e^{-c^2}$ to ensure that $0 \leq r(c) \leq 1$.
We summarise the main differences in the following 6 points.
\begin{enumerate}
	\item
	The solution of simple gradient descent highly depends on the parametrisation.	
	\item Even our stable polar parametrisation gets trapped at the origin of the Bloch ball due to the vanishing
	gradient $\underline{g}(c,\theta) = (2 e^{-c^2} c \cos \theta,  [e^{-c^2} -1] \sin \theta)$.
	This gradient vanishes when\ $c=0$ or equivalently when $r=0$ as shown in Fig.~\ref{analyitc-fig}(b).
	\item
	In contrast, natural gradient descent is independent of the parametrisation:
	Besides our analytical solution, we numerically simulate
	the polar parametrisation $\overrightarrow{r}=(r(c),\theta) $ in Fig.~\ref{analyitc-fig}(a)
	(blue line). We compare the numerical evolution to our analytical solution
	and indeed see a perfect match in Fig.~\ref{analyitc-fig}(c) confirming the
	independence from parametrisations.
	\item The eigenvalues of the metric tensor
	are $(\frac{1}{1-\lvert \overrightarrow{r} \rvert^2}, 1)$,
	which are singular for the limiting case of pure states, i.e.,  when  $\lvert \overrightarrow{r} \rvert=1$.
	This nicely demonstrates that when approaching the optimal solution -- which is a pure state --
	small steps in coordinate space result in increasingly large `jumps' in state space.
	The inverse of the metric tensor corrects for this effect and slows down the evolution
	when approaching pure states. In contrast, the simple gradient is constant
	via $\underline{g}(y,z) = (0,1)$. We remark that the above discussed singular behaviour for approaching pure states
	is expected to be completely general.
	\item
	The simple gradient descent evolution seems to converge to the solution in only inverse polynomial order
	in  Fig.~\ref{analyitc-fig}(d).
	In contrast, the natural gradient evolution has a fundamentally
	improved exponential convergence rate in Eq.~\eqref{descent-solution},
	and we illustrate this in Fig.~\ref{analyitc-fig}(d).
	\item
	Finally, we emphasise that the metric tensor and the natural gradient
	approach become increasingly important when considering non-unitary evolutions.
	In particular, it is straightforward to see that when restricting the evolution to pure states
	via the polar parametrisation $\overrightarrow{r}= (r=1, \theta)$, then the metric tensor becomes constant and
	therefore irrelevant.
\end{enumerate}

\section{Proofs \label{app:proofs}}

The metric tensor $\matra$ has been related to the classical Fisher information
in~\cite{facchi2010classical, petz_book_chapter, petz_geometry}
as $\matra = \cf/4$, if the state vector is isomorphic to a classical probability distribution $p( n |\underline{\theta} )$.
In this case the state vector
$\psi(\underline{\theta}) = \sum_n \sqrt{p( n |\underline{\theta} )} \, |n\rangle$
contains no phase information.
Stokes et al.~concluded~\cite{quantumnatgrad} that the resulting update rule in Eq.~\eqref{imagtimeUpdate}
is identical to the natural gradient optimisation well-known in the context of machine learning
\cite{amari1997neural,goodfellow2016deep,amari2000adaptive}
-- which  uses the classical Fisher information matrix $\cf$ as a metric tensor in Eq.~\eqref{naturalgradEvo}.
Let us now provide a concise proof of a slightly different and more general statement.

\begin{theorem} \label{class-fisher}
	If the state vector is isomorphic to a
	classical probability distribution (in the computational basis) as $\psi(\underline{\theta}) = \sum_n \sqrt{p( n |\underline{\theta} )} \, |n\rangle$,
	then the quantum and classical Fisher information matrices are equivalent via $[\qf]_{kl} = [\cf]_{kl}$.
	The optimal measurement basis for obtaining the quantum Fisher information in Sec.~\ref{mainres} is just
	the standard computational basis $\{ |n \rangle \}$ and one can establish a series of equations
	\begin{equation*}
		[\qf]_{kl} =  4 [\matra]_{kl} = 4 \langle \partial_k \psi | \partial_l \psi \rangle = [\cf]_{kl}  =  \sum_n p( n |\underline{\theta} ) \bigg(\frac{\partial\mathrm{ln}    ~ p( n |\underline{\theta} )}{\partial \theta_k} \bigg)
		\bigg(\frac{\partial\mathrm{ln}    ~ p( n |\underline{\theta} )}{\partial \theta_l} \bigg).
	\end{equation*}
	Here the classical Fisher information $[\cf]_{kl}$
	depends on the classical probability distribution
	$p( n |\theta ) = |\langle \psi(\underline{\theta}) | n \rangle|^2$
	that is obtained vy measuring the quantum state in the standard measurement basis.
\end{theorem}

\begin{proof}
	We start with the explicit expression from Remark~\ref{remark1} and we drop the second
	term to obtain
	\begin{equation*}
		[\qf]_{kl} = 4 [\matra]_{kl} = 4 \langle \partial_k \psi | \partial_l \psi \rangle,
	\end{equation*}
	which expression is valid for state vectors whose global phase does not evolve under the variation
	of the parameters. Let us now expand the partial derivatives as
	\begin{equation*}
		\partial_k \psi  =  \partial_k \sum_n \sqrt{p( n |\underline{\theta} )} \, |n\rangle
		=
		\sum_n \frac{\partial_k p( n |\underline{\theta} ) }{ 2 \sqrt{p( n |\underline{\theta} )}  }\, |n\rangle.
	\end{equation*}
	Let us now compute the scalar products as
	\begin{equation*}
		4 \langle \partial_k \psi | \partial_l \psi \rangle
		=
		4 \sum_n
		\frac{\partial_k p( n |\underline{\theta} ) }{ 2 \sqrt{p( n |\underline{\theta} )}  } 
		\frac{\partial_l p( n |\underline{\theta} ) }{ 2 \sqrt{p( n |\underline{\theta} )}  } 
		=
		\sum_n
		\frac{\partial_k p( n |\underline{\theta} )  \partial_l p( n |\underline{\theta} ) }{  p( n |\underline{\theta} )} .
	\end{equation*}
	We now use the equality of derivative functions $ \tfrac{\mathrm{d} f(x)}{\mathrm{d} x} /f(x) = \frac{\mathrm{d} \ln(f(x)) }{\mathrm{d} x}$
	as
	\begin{equation*}
		4 \langle \partial_k \psi | \partial_l \psi \rangle
		=
		\sum_n
		p( n |\underline{\theta})
		\bigg( \frac{ \partial \mathrm{ln} ~ p( n |\underline{\theta} ) } {\partial \theta_k} \bigg)
		\bigg( \frac{ \partial \mathrm{ln} ~ p( n |\underline{\theta} ) } {\partial \theta_l} \bigg)
		\equiv
		[\cf]_{kl},
	\end{equation*}
	where the last equality confirms that indeed, we get the expression for the classical Fisher information
	and the optimal measurement basis is the standard basis $\{ |n \rangle \}$.
\end{proof}

\begin{theorem}\label{Fourier_series_theorem}
	Let us define $L$-local (unitary or CPTP) transformations as mappings
	$\Phi(\theta)$ over density matrices that act non-trivially on only $L$ qubits.
	A circuit composed of $\nu$ $L$-local unitary gates
	can be expanded into $\mathcal{O}(2^{2 \nu L})$ Fourier components while a similar circuit composed of
	non-unitary gates can be decomposed into $\mathcal{O}(2^{4 \nu L})$ Fourier components.
	It follows that objective functions of the form $E(\underline{\theta}) = \tr[\rho(\underline{\theta})  \mathcal{H}]$
	can be expressed as Fourier series of $\mathcal{O}(2^{2 \nu L})$  or $\mathcal{O}(2^{4 \nu L})$ terms, respectively.
\end{theorem}

\begin{proof}
	We define the full circuit as
	\begin{equation*}
		\Phi_C(\theta) := \prod_{k=1}^{\nu} \Phi_k (\theta_k).
	\end{equation*}
	Since every $\Phi_k (\theta_k)$ can be expanded into $\mathcal{O}(2^{2L})$ terms
	via Lemma~\ref{lemma_dimensions}, it immediately follows that the above product
	can be expanded into a sum of $\mathcal{O}(2^{2 \nu L})$ Fourier components.

	One can similarly argue about general CPTP maps using Lemma~\ref{lemma_dimensions}
	and conclude that such circuits can be expanded into $\mathcal{O}(2^{4 \nu L})$
	Fourier components.
\end{proof}

\begin{lemma}\label{lemma_dimensions}
	Continuous, parametrised $L$-local unitary transformations from
	Theorem~\ref{Fourier_series_theorem}
	can be expanded into
	a Fourier series of $\mathcal{O}(2^{2L})$ terms while CPTP transformations
	can be expanded into  $\mathcal{O}(2^{4L})$ terms. We emphasize that
	despite our proof techniques use purifications, one can efficiently implement
	an $L$-local non-unitary transformation on a quantum computer of $N$ qubits
	without the need for ancilla qubits as we show in Sec.~\ref{effective-cptp-map} (\spl).
\end{lemma}
\begin{proof}
	Let us first prove the unitary case. We denote an arbitrary, parametrised $L$-local unitary as
	$U(\theta) \in SU(2^L)$ and expand it into its eigenbasis as
	\begin{equation*}
		U(\theta) = \sum_{k = 1}^{S_1} e^{-i \theta E_k} P_k,
	\end{equation*}
	where $E_k$ are eigenvalues and $P_k$ are projectors onto their eigenspaces.
	We denote the number of distinct eigenvalues as $S_1 \leq 2^L$ which is
	upper bounded by the dimensionality.
	We now assume density matrices of rank-1 as $\rho:= |\psi \rangle \langle \psi|$.
	Our $L$-local unitary mapping can be defined as
	$\Phi(\theta) \rho := U(\theta) \rho U^\dagger(\theta)$ and it decomposes into
	the expected number of terms via
	\begin{equation*}
		\Phi(\theta) = \sum_{k,l = 1}^{S} e^{-i \theta (E_k-E_l)} \mathcal{P}_{kl},
		\quad \quad \text{with} \quad \quad
		\mathcal{P}_{kl} \rho := P_k \rho P^\dagger_l.
	\end{equation*}
	Indeed this is a Fourier series of $\mathcal{O}(S^2) = \mathcal{O}(2^{2L})$ terms.

	We now consider $L$-local non-unitary, general CPTP maps. Recall that any CPTP
	map can be represented as a unitary transformation in an enlarged Hilbert
	space due to the so-called Stinespring dilation \cite{stinespring1955positive}.
	In particular, let us represent our quantum state $\rho$ of $N$ qubits as
	a the purification $\tilde{\psi}$ in a larger Hilbert space.
	We indeed by definition recover $\rho =: \tr_{anc}[\tilde{\psi}]$,
	where $\tr_{anc}$ denotes the partial trace over the ancillary Hilbert space.
	We now use that any $L$-local CPTP map over $\rho$ can be written as
	a unitary transformation over the purification $\tilde{\psi}$ and this unitary
	need at most be $2L$ local. (An explicit construction could use $L$ ancillary qubits coupled to every
	tuple of $L$ qubits.)
	Using our proof from the previous paragraph, but now for $2L$-local
	unitary maps, we conclude that $L$-local CPTP maps can be decomposed into 
	a Fourier series of $\mathcal{O}(2^{4L})$ terms.

	We emphasise that one can effectively implement such non-unitary
	maps in practice without the need for ancillary qubits as shown in Sec.~\ref{effective-cptp-map} (\spl).
\end{proof}

\section{Possible construction of effective CPTP maps \label{effective-cptp-map}}
In this section we give a possible example of constructing $K$-local CPTP maps
without using ancillary qubits.
In this example
one only needs to implement unitary transformations randomly at every
execution of the circuit. We illustrate this on the example of an arbitrary
$1$-local CPTP map. An arbitrary single qubit state $\rho$ can be uniquely
represented as a point in the Bloch ball, which we parametrise in terms of
the usual polar coordinates $(r, \theta, \phi)$.
Unitary transformations applied to
$\rho$ rotate the Bloch sphere, hence transforming the angles $(\theta, \phi)$.
Decreasing the length of the Bloch vector $r$ can be performed by
effectively applying the depolarising channel, which we show via
its effect on any observable $\mathcal{H}$
\begin{equation}
	\tr[\rho' \mathcal{H}] = (1-p) \tr[\rho \mathcal{H}] 
	+ \frac{p}{3} \left\{ \tr[X \rho X \, \mathcal{H}] + \tr[Y \rho Y \, \mathcal{H}] + \tr[Z \rho Z \, \mathcal{H}]\right\}.
\end{equation}
Here, one only needs to implement unitary reflections via the Pauli operators $X, Y$ and $Z$ with probability $p$ and
sum together the resulting expectation values. Increasing the length
of the Pauli vector $r$ can be done similarly, but by applying `negative probabilities' (when $\rho \neq \mathrm{Id}$).
In particular, one repeats the previous protocol of applying unitary reflections with the same probabilities,
and changes the sign of the probabilities when summing up the resulting expectation values.
The above construction provides a full parametrisation and every $1$-local CPTP map can be realised
this way.

We finally remark that our example is an instance of the following general constructions of 
	parametric CPTP maps as
\begin{equation*}
	\rho(\underline{p}) = \sum_{k=1}^{N_p} p_k U_k | \psi \rangle \langle \psi | U_k^\dagger,
\end{equation*}
for an input pure state $| \psi \rangle$ and $N_p$ unitary circuits $\{U_k\}$ with probabilities $\sum_k p_k = 1$.
The resulting expectation values can be estimated via
\begin{equation*}
	\tr[ \rho(\underline{p}) \mathcal{H} ] = \sum_{k=1}^{N_p} p_k  \tr[ U_k | \psi \rangle \langle \psi | U_k^\dagger \mathcal{H} ],
\end{equation*}
e.g., a stochastic sampling from a series of circuits $U_k$.

\section{Derivation of Result~\ref{result3}}

\begin{figure*}[tb]
	\begin{centering}
		\includegraphics[width=0.6\textwidth]{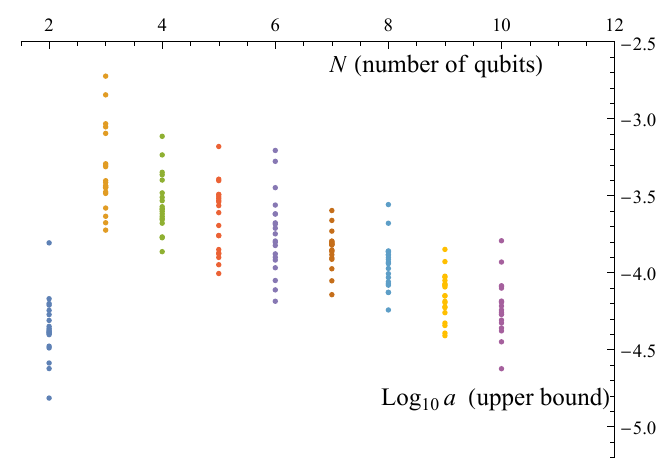}
		\caption{
			Upper bounds $a:= \max_{k} [\sum_{m=1}^d |\partial_k \lambda_m|^2]$
			as a function of the number of qubits computed for each $N$ at $20$ sets of random parameters
			of an ansatz circuits as in Fig.~\ref{circ} with $15$ repeating layers. For each set of parameters we compute 
			the norm of the derivatives of the eigenvalue vector as $\sum_{m=1}^d |\partial_k \lambda_m|^2$ for
			each parameter $\theta_k$
			and only plot the maximum over $k$ as our upper bound $a$.	2-qubit gates in the ansatz circuit
			undergo 2-qubit depolarising noise with probability $p_{err}$ while single qubit gates undergo
			depolarising noise with probability $0.1 p_{err}$. We set $p_{err}$ such that the number of expected
			errors in the circuit is constant $1$ for all system sizes.
			Classically computing $a$ is more efficient than computing the QFI and this allows us to consider significantly
			larger system sizes and circuit depths than in Fig.~\ref{approximations}.
		}
		\label{fig:aterms}
	\end{centering}
\end{figure*}

\begin{definition}\label{definition:bterm}
	Given an arbitrary quantum state in terms of its spectral decomposition as
	$\rho := \lambda |\psi \rangle \langle \psi | +  \sum_{m=2}^d \lambda_m |\psi_m \rangle \langle \psi_m |$,
	we consider column vectors of the symmetric logarithmic derivative from Eq.~\eqref{symmlogdermulti} as 
	\begin{equation*}
	  L_k |\psi_m \rangle = \sum_{n=1}^{d}  \langle \psi_n | L_k | \psi_m \rangle \, |\psi_n \rangle =:
	  \frac{\partial_k \lambda_m}{\lambda_m}  |\psi_m \rangle  + |\phi_{mk}\rangle,
	\end{equation*}
	under some continuous parametrisation $\rho(\underline{\theta})$.
	Above we define the off-diagonal part of the column vectors as $|\phi_{mk}\rangle$,
	and we derive an upper bound on their norm below as
	\begin{equation*}
	\lVert \phi_{mk} \rVert^2 \leq b := \max_{m,k}  \rVert \partial_{k} \psi_m  \lVert^2.
	\end{equation*}
	We discuss below that $b$ is constant bounded in case of typical ansatz circuits and, e.g., if the parametrisation of the quantum state
	$\rho(\underline{\theta})$ is generated by Pauli operators (most common universal construction)
	then $b\leq 1$. 
	Furthermore, we define the upper bound on the norm of the vector of derivatives of the eigenvalues	as
	\begin{equation*}
		|\sum_{m=1}^d (\partial_k \lambda_m) (\partial_l \lambda_m)| \leq 
		a:= \max_{k} [\sum_{m=1}^d |\partial_k \lambda_m|^2].
	\end{equation*}
	We derive 3 scenarios below. a) In the limit of global depolarising noise
	$\rho := \lambda |\psi \rangle \langle \psi | + \tfrac{1-\lambda}{d} \mathrm{Id}$
	we obtain $a=0$. b) For unitary parametrisations of arbitrary mixed quantum states as
	$U(\underline{\theta}) \rho U(\underline{\theta})^\dagger$
	we similarly obtain $a=0$. c) If the parametrisation corresponds to
	quantum circuits of local gates affected by local errors of small probability as in case of typical
	error models in near-term quantum devices,
	it can be expected that $a \ll 1$ via  ref.~\cite{koczor2021dominant}.
	In Fig.~\ref{fig:aterms} we provide comprehensive numerical evidence that indeed $a \ll 1$ and
	that it decreases as we increase the system size -- since the error eigenvalues $\lambda_m$ approach close to maximum entropy
	as global depolarising noise.
\end{definition}

\begin{proof}
	
	A general expression for the matrix elements $L_k$ can be found in, e.g., reference \cite{qfi1} as
	\begin{equation} \label{matrelems}
		\langle \psi_m | L_l |\psi_n \rangle =  \delta_{mn} \frac{\partial_l \lambda_m}{\lambda_m} + \frac{2(\lambda_n - \lambda_m)}{\lambda_n + \lambda_m} \langle \psi_m |\partial_l \psi_n \rangle,
	\end{equation}
	where $\lambda_m$ are eigenvalues of the density matrix and we can use the above formula
	to explicitly compute the norm of the (off-diagonal) column vectors as
	\begin{equation*}
		\lVert \phi_{mk} \rVert^2 :=   \sum_{n\neq m}^{d}  |\langle \psi_n | L_k | \psi_m \rangle |^2
		= 
	  \sum_{n \neq m}^{d}  
		|\frac{2(\lambda_n - \lambda_m)}{\lambda_n + \lambda_m}|^2 |\langle \psi_n |\partial_{k} \psi_m \rangle|^2
		\leq  \rVert \partial_{k} \psi_m  \lVert^2
	\end{equation*}
	where we have used that $	|\frac{2(\lambda_n - \lambda_m)}{\lambda_n + \lambda_m}|^2 \leq 1$.
	
	In most typical quantum circuits the parametrisation corresponds to quantum gates that are generated by some (effective) Hamiltonian
	$\mathcal{H}$ in $U(\theta) = e^{-i \theta \mathcal{H}}$.
	The derivatives as $\rVert \partial_{k} \psi_m  \lVert$ are therefore bounded via the norm
	$\lVert \frac{ \partial U(\theta)} {\partial \theta} \rVert_{\infty}
	=
	\lVert -i \mathcal{H} U(\theta) \rVert_{\infty} \leq \lVert \mathcal{H} \rVert_{\infty}	$.
	Here the norm of the generator $\mathcal{H}$ is independent of the system size when the
	generator is local (only acts on 1 or 2 qubits in typical quantum gates). It follows that under these assumptions
	$\rVert \partial_{k} \psi_m  \lVert$ is constant bounded and therefore $b$ is also constant bounded.
	For example, in quantum circuits that consist of Pauli gates as $\mathcal{H} = P$ for some Pauli string $P \in \{\mathrm{Id},\sigma_x,\sigma_y,\sigma_z\}^{\otimes N}$,
	we obtain $\lVert \mathcal{H} \rVert_{\infty}  = 1 $ and therefore $b\leq 1$. 
	These arguments can straightforwardly be extended to local noisy gates as Markovian
	processes via $\Phi(\theta) = e^{-i \theta \mathcal{L}}$, where $\mathcal{L}$ is a local superoperator
	that is the generator of the non-unitary dynamics. Note that the local noise assumption is expected to apply
	to typical error models in near-term quantum devices.
	
	We can also straightforwardly extend these arguments to non-local parametrisations too. For example, in the
	case of the variational Hamiltonian ansatz, as in the case of QAOA \cite{farhi2014quantum}, the parametrisation
	is similarly of the form $U(\theta) = e^{-i \theta \mathcal{H}}$, but here the generator 
	$\mathcal{H} = \sum_{k=1}^h c_k P_k$ is the non-local problem Hamiltonian that consists of some Pauli strings $P_k$. Given the usual polynomial
	growth in the numbers of qubits as $h \in \mathrm{poly}(N)$ our upper bound scales as $b \in \mathrm{poly}(N)$. 
	In such a scenario our error bound in Result~\ref{result3} is guaranteed decrease with the system size
	when $p_{max}$ decreases faster than the growth rate of $h$. Indeed, in the limit of depolarising noise $p_{max}$ 
	decreases exponentially while $b \in \mathrm{poly}(N)$.
	
	The upper bound
		\begin{equation*}
		|\sum_{m=1}^d (\partial_k \lambda_m) (\partial_l \lambda_m)| \leq 
		a:= \max_{k} [\sum_{m=1}^d |\partial_k \lambda_m|^2]
	\end{equation*}
	is a straightforward consequence of the Cauchy-Swartz inequality.
	
	a) In the case of global depolarising noise as $\rho := \lambda |\psi \rangle \langle \psi | + \tfrac{1-\lambda}{d} \mathrm{Id}$,
	the parametrisation is independent of the noise model and therefore $(\partial_k \lambda_m) = 0$
	for all $m,k$. It follows that $a=0$.
	
	b) We can apply the same argument in case of unitary parametrisations
	of arbitrary mixed quantum states as $U(\underline{\theta}) \rho U(\underline{\theta})^\dagger$
	and obtain $a=0$. 
	
	c) Ref.~\cite{koczor2020exponential,koczor2021dominant,huggins2020virtual} established that in typical near-term quantum devices
	the dominant eigenvector generally $| \psi \rangle \langle \psi|$ approximately commutes with the noise operators.
	Furthermore, the limit of global depolarising noise has been shown to be a good approximation of experimental
	noise models \cite{PhysRevE.104.035309,arute2019quantum,koczor2021dominant, koczor2020exponential, huggins2020virtual}
	which implies $a\approx 0$, and this approximation enhances as we increase the system size.
	For these reasons we can expect that the eigenvalues of $\rho$ are approximately
	constant under small variations of the parameters and therefore $a \ll 1$. As such, we numerically
	validate that indeed the derivatives $a_k:=\sum_{m=1}^d |\partial_k \lambda_m|^2]$ are small in practice and
	decrease when we increase the system size in Fig.~\ref{fig:aterms}.
\end{proof}

\begin{lemma}\label{lemma:sqterm}
	Given any quantum state in terms of its spectral decomposition
	$\rho := \lambda |\psi \rangle \langle \psi | +  \sum_{m=2}^d \lambda_m |\psi_m \rangle \langle \psi_m |$,
	we obtain the following approximation in terms of the dominant eigenvector $|\psi_1 \rangle \equiv |\psi \rangle$ as
	\begin{equation*}
		\tr[\rho \rho L_k L_l ] = \lambda^2 \tr[ |\psi \rangle \langle \psi | L_k L_l] + E_1,
	\end{equation*}
	where the error term is upper bounded as $|E_1| \leq   a + b \lambda_2 (1-\lambda)$ which scales
	with the dominant error eigenvalue as $E_1 \in a + \mathcal{O}(\lambda_2)$ up to an additive error $a$
	that is small in practice under assumptions from Definition~\ref{definition:bterm}.
\end{lemma}
\begin{proof}
	We first compute the square of the density matrix and use the linearity of the trace operation as
	\begin{equation*}
		\tr[\rho \rho L_k L_l ] =
		\lambda^2 \tr[ |\psi \rangle \langle \psi | L_k L_l] +  \sum_{m=2}^d \lambda_m^2 \,  \tr[|\psi_m \rangle \langle \psi_m | L_k L_l] 
	\end{equation*}
	and we denote the second term above as $E_1$ and upper bound it as
	\begin{equation*}
	E_1 =   \sum_{m=2}^d \lambda_m^2  \tr[|\psi_m \rangle \langle \psi_m | L_k L_l] 
	=   \sum_{m=2}^d \lambda_m^2    \langle \psi_m | L_k L_l |\psi_m \rangle 
	= \sum_{m=2}^d \lambda_m^2
	[ \frac{\partial_k \lambda_m}{\lambda_m}  \frac{\partial_l \lambda_m}{\lambda_m} 
	 +  \langle \phi_{mk} |  \phi_{ml} \rangle  ] ,
  	\end{equation*}
    where in the last equation we used Definition~\ref{definition:bterm}. We can therefore upper bound the error as
  	\begin{equation*}
  	|E_1| \leq  
  	 |\sum_{m=2}^d (\partial_k \lambda_m) (\partial_l \lambda_m)|
  	+
  	 \sum_{m=2}^d \lambda_m^2  |\langle \phi_{mk} |  \phi_{ml} \rangle |
  	 \leq a + b \lambda_2 (1-\lambda)
    \end{equation*}
	where we used Cauchy-Schwartz inequality in combination with our bounds
	from Definition~\ref{definition:bterm} as $|\langle \phi_{mk} |  \phi_{ml} \rangle |\leq b$
	and we evaluated the sum $\sum_{m=2}^d \lambda_m = 1-\lambda$.
\end{proof}

\begin{lemma} \label{lemma:depol}
	If the quantum state undergoes global depolarising noise as $\rho := \lambda |\psi \rangle \langle \psi | + \tfrac{1-\lambda}{d} \mathrm{Id}$,
	we can approximate the term from Lemma~\ref{lemma:sqterm} in terms of the density matrix instead of the dominant eigenvector as
	\begin{equation*}
		\tr[\rho \rho L_k L_l ] = \lambda \tr[ \rho L_k L_l] + E_{depol,1}.
	\end{equation*}
	Here the error term is upper bounded as $|E_{depol,1}| \leq b [
	\tfrac{(1-\lambda)^2}{d^2} +  \lambda \tfrac{1-\lambda}{d} ]$
	and therefore scales with the dominant error eigenvalue as $E_{depol,1} \in \mathcal{O}( 1/d )$
	which scales with the dimensionality $d$.
	Here the multiplicative factor $b$ is defined in Definition~\ref{definition:bterm} and is shown to be
	constant bounded (or at most polynomially growing) in case of parametrised quantum circuits.
\end{lemma}
\begin{proof}
	Let us first compute the square of the density operator $\rho^2$ as
	\begin{equation*}
		\rho^2 = [ \lambda |\psi \rangle \langle \psi | + \tfrac{1-\lambda}{d} \mathrm{Id} ]^2
		=
		\lambda^2 |\psi \rangle \langle \psi | + 2 \lambda \tfrac{1-\lambda}{d}  |\psi \rangle \langle \psi | 
		+\tfrac{(1-\lambda)^2}{d^2} \mathrm{Id}
		=
		 \lambda [ \lambda |\psi \rangle \langle \psi | + \tfrac{1-\lambda}{d} \mathrm{Id}]
		+ R,
	\end{equation*}
where the residual operator is $R:=  [\tfrac{(1-\lambda)^2}{d^2} - \lambda \tfrac{1-\lambda}{d}]
 \mathrm{Id}  + 2 \lambda \tfrac{1-\lambda}{d}  |\psi \rangle \langle \psi | $.
\\[5mm]
For ease of notation, we re-write the above expression in terms of the ideal term plus the
residual operator as
	\begin{equation*}
	\tr[\rho \rho L_k L_l ] = \lambda \tr[\rho L_k L_l ] + \tr[R L_k L_l ].
	\end{equation*}
Using the above simplified form we can compute the error term in terms of the residual operator as
\begin{equation}\label{e1def_depol}
	E_{depol,1} :=  \tr[R L_k L_l ] = [\tfrac{(1-\lambda)^2}{d^2} - \lambda \tfrac{1-\lambda}{d}]
	\tr[L_k L_l ]
	+
	2 \lambda \tfrac{1-\lambda}{d} 
		\langle \psi | L_k L_l  |\psi \rangle .
\end{equation}
We compute the first term from the above equation as
\begin{equation*}
	\tr[L_k L_l ] = \sum_{m=1}^d \langle \psi_m | L_k L_l  |\psi_m \rangle
	=
	\sum_{m=1}^d \langle \phi_{mk}   |\phi_{ml} \rangle
	=
	\langle \phi_{1k}   |\phi_{1l} \rangle
\end{equation*}
where we have used that under global depolarising noise only the first row and column of $L_k$
are non-zero via the matrix elements
\begin{equation}
	\langle \psi_m | L_l |\psi_n \rangle =  \delta_{mn} \frac{\partial_l \lambda_m}{e_m} + \frac{2(\lambda_n - \lambda_m)}{\lambda_n + \lambda_m} \langle \psi_m |\partial_l \psi_n \rangle,
\end{equation}
as	 the matrix elements $\langle \psi_m | L_l |\psi_n \rangle = 0$ are zero for all $m,n \geq 2$ and when $m=n=1$
since the eigenvalues of the density matrix are identical $\lambda_m = 1/d$ for all $m\geq 2$ and
derivatives of the eigenvalues are zero $|\partial_l \lambda_m|= 0$ as the noise model (global depolarising noise) is
independent of the parametrisation. We can finally upper bound the term via Definition~\ref{definition:bterm} as
\begin{equation*}
	|\tr[L_k L_l ]|
	=
	|\langle \phi_{1k}   |\phi_{1l} \rangle|
	\leq \lVert \phi_{1k} \rVert\lVert \phi_{1l} \rVert \leq b.
\end{equation*}

We can similarly upper bound  the second term in Eq.~\eqref{e1def_depol} as
$|\langle \psi | L_k L_l  |\psi \rangle| \leq b$ via Definition~\ref{definition:bterm}.

We finally conclude that
\begin{equation*}
	\tr[\rho \rho L_k L_l ] = \lambda \tr[\rho L_k L_l ] + E_{depol,1},
\end{equation*}
where the error term is upper bounded as $|E_{depol,1}| \leq b [
\tfrac{(1-\lambda)^2}{d^2} +   \lambda \tfrac{1-\lambda}{d} ]$
scales with the dominant error eigenvalue as $E_{depol,1} \in \mathcal{O}( \tfrac{1-\lambda}{d} )$.
\end{proof}

\begin{lemma}\label{lemma_errorterm}
	Given any quantum state in terms of its spectral decomposition
	$\rho := \lambda |\psi \rangle \langle \psi | + \sum_{m=2}^d \lambda_m |\psi_m \rangle \langle \psi_m |$
	with using the notation $|\psi_1 \rangle \equiv |\psi \rangle$,
	we can upper bound the following error term as
\begin{equation*}
	|E_2| := |\tr[\rho L_k \rho L_l ] |
	\leq
		a + b (1+\lambda)\lambda_2
\end{equation*}
which depends on the largest error probability up to an additive error $a$ as
$E_2 = a + \mathcal{O}(\lambda_2)$. The additive error
expresses how much the eigenvalues can change under a variation of the parameters as discussed in Definition~\ref{definition:bterm}.
\end{lemma}
\begin{proof}
	Let us introduce the notation $\rho =  \lambda |\psi \rangle \langle \psi | + \rho_{err}$ with $\rho_{err}:=\sum_{m=2}^d \lambda_m  |\psi_m \rangle \langle \psi_m | $ and substitute it into our expression as
		\begin{align} \label{error_terms}
		&\tr[\rho L_k \rho L_l ] = \lambda^2 \langle \psi | L_k |\psi \rangle\langle \psi | L_l |\psi \rangle
		+
		\lambda[ \langle \psi | L_k \rho_{err} L_l |\psi \rangle +  \langle \psi | L_l \rho_{err} L_k |\psi \rangle]
		+ \tr[\rho_{err} L_k \rho_{err} L_l ],
	\end{align}
where we have evaluated the trace operation in the case of the pure state $|\psi_1 \rangle$.
Using the explicit expression for matrix elements from Eq.~\eqref{matrelems},
the first term above is given by the derivatives of the dominant eigenvalue as
$$\lambda^2 \langle \psi | L_k |\psi \rangle\langle \psi | L_l |\psi \rangle = (\partial_k \lambda )(\partial_l \lambda).$$

The second term in Eq.~\eqref{error_terms} is bounded as
\begin{equation*}
	\lambda | \langle \psi | L_k \rho_{err} L_l |\psi \rangle|
	= \lambda |\langle \phi_{1k} | \rho_{err} |\phi_{1l} \rangle|
	\leq
	\lambda \lVert \rho_{err} \rVert_{\infty} \lVert \phi_{1k} \rVert \lVert \phi_{1l} \rVert
	\leq 
	\lambda b \lambda_2,
\end{equation*}
here the infinity norm of $\rho_{err}$ is given by its largest eigenvalue as
$\lVert \rho_{err} \rVert_{\infty} = \lambda_2$
and we have derived the upper bound $\lVert \phi_{ml} \rVert \leq \sqrt{b}$ on the norm of the column vectors in Definition~\ref{definition:bterm}.
\\[5mm]
The third term is given as
\begin{equation*}
	\tr[\rho_{err} L_k \rho_{err} L_l ] = \sum_{m=2}^d \lambda_m \langle \psi_m | L_k \rho_{err} L_l |\psi_m \rangle 
=
\sum_{m=2}^d  (\partial_k \lambda_m)(\partial_l \lambda_m)
+
\sum_{m=2}^d \lambda_m   \langle \phi_{km} | \rho_{err}  |\phi_{lm} \rangle 
\end{equation*}
where we  can upper bound the second term as
\begin{equation*}
	|\sum_{m=2}^d \lambda_m   \langle \phi_{km} | \rho_{err}  |\phi_{lm} \rangle|
	\leq 
	\sum_{m=2}^d \lambda_m   |\langle \phi_{km} | \rho_{err}  |\phi_{lm} \rangle|
	\leq 
	b (1-\lambda)\lambda_2
\end{equation*}
using the inequality from above as $| \langle \phi_{km} | \rho_{err}  |\phi_{lm} \rangle  | \leq  b \lambda_2$ 
and we have evaluated the sum $\sum_{m=2}^d \lambda_m = 1-\lambda$.

Recollecting all terms and using the bound $|\sum_{m=1}^d  (\partial_k \lambda_m)(\partial_l \lambda_m)| \leq a $
from Definition~\ref{definition:bterm} we find that
\begin{equation*}
	|E_2| = |\tr[\rho L_k \rho L_l ] |
	\leq
	a + 2 b \lambda  \lambda_2 	+ b (1-\lambda)\lambda_2
	=
	a + b (1+\lambda)\lambda_2
\end{equation*}
and therefore the error term scales with the dominant error eigenvalue up to an additive
error $a$ that expresses how much the eigenvalues can change under the
variation of any parameters via Definition~\ref{definition:bterm}. 

\end{proof}

\begin{theorem}\label{theo1}
	The Hilbert-Schmidt metric tensor of any noisy quantum state $\rho$ approximates the quantum Fisher information
	of the dominant eigenvector $|\psi \rangle$ of the same quantum state up to an error $E$ as
	\begin{equation*}
		\tr[(\partial_k \rho)(\partial_l \rho)]
		=
			 \frac{\lambda^2}{2} [\qf(|\psi\rangle\langle\psi|)]_{kl} + E,
			 \quad \quad \text{where} \quad \quad |E|	\leq   a +  b \lambda_2 .
	\end{equation*}
	We discuss in Definition~\ref{definition:bterm} that $b$ is constant bounded in usual
	quantum circuits and therefore $E$ scales with the dominant error eigenvalue as $|E| = a  + \mathcal{O}(\lambda_2)$
	up to an additive error $a$ which expresses how much the eigenvalues can change under the
	variation of any parameters via Definition~\ref{definition:bterm}.
	Furthermore, we show in Definition~\ref{definition:bterm} that in the limit of global depolarising noise $a = 0$
	and we therefore obtain the simplified expressions
	\begin{equation*}
		\tr[(\partial_k \rho)(\partial_l \rho)] =
		\frac{\lambda^2}{2} [\qf(|\psi\rangle\langle\psi|)]_{kl}	+ E_{depol}
		 \quad \quad \text{where} \quad \quad |E_{depol}|	\leq    b (1-\lambda)/d,
	\end{equation*}
	which guarantees an approximation error $E_{depol} \in \mathcal{O}(1/d)$ that scales inversely with the
	dimensionality.
	Furthermore, under global depolarising noise, we can also approximate the quantum Fisher information
	of any mixed quantum state (and not just its dominant eigenvector) as
	\begin{equation*}
		\tr[(\partial_k \rho)(\partial_l \rho)] =	\frac{\lambda}{2} [\qf(\rho)]_{kl} + E'_{depol},
				 \quad \quad \text{where} \quad \quad
		|E'_{depol}| \leq   b [	\tfrac{(1-\lambda)^2}{d^2} +  \lambda \tfrac{1-\lambda}{d} ]/2 +  b(1-\lambda)(1+\lambda)/(2d).
	\end{equation*}
	Here the approximation error $E'_{depol} \in \mathcal{O}[(1-\lambda)/d]$ similarly scales inversely with the
	dimensionality.
\end{theorem}

\begin{proof}
Let us use the defining expression of the symmetric logarithmic derivative from Eq.~\eqref{symmlogdermulti} as
	\begin{equation*}
		4 \tr[(\partial_k \rho)(\partial_l \rho)] =
	\tr[
	( L_k \rho + \rho L_k) \, (L_l \rho + \rho L_l)]
	\end{equation*}
	The expression in the right-hand side can
	be expanded into four terms  as
	\begin{align*}
		4	\tr[(\partial_k \rho)(\partial_l \rho)] =
			\tr[L_k \rho L_l \rho] +  \tr[L_k \rho \rho L_l]
		+
		\tr[\rho L_k L_l \rho] + \tr[\rho L_k \rho L_l ].
	\end{align*}
	Using the the cyclic property of the trace operation results in
	$ \tr[\rho L_k L_l \rho] =  \tr[\rho \rho L_k L_l ]$
	and in $\tr[L_k \rho L_l \rho] =  \tr[\rho L_k \rho L_l ]$.
	This yields the simplified form
	\begin{align*}
		4	\tr[(\partial_k \rho)(\partial_l \rho)] =&
		\tr[\rho \rho L_l L_k] + \tr[\rho \rho L_k L_l ]  + 2 \tr[\rho L_k \rho L_l ] \\
		=&  \lambda^2 \tr[|\psi \rangle \langle \psi |  (L_k L_l  +  L_l L_k) ]  + 2 (E_1 + E_2),
	\end{align*}
	where the second equality is due to Lemmas~\ref{lemma:sqterm} and \ref{lemma_errorterm}.
	
	Let us now recall that by definition (see Eq.~\eqref{qfi-mat-def})
	the quantum Fisher information is given by the expectation value
	$2 [\qf]_{kl} := \tr[\rho(\underline{\theta}) \, (L_k L_l + L_l L_k) ]$
	and therefore we obtain the result
	\begin{equation}
	 \tr[(\partial_k \rho)(\partial_l \rho)] =
		 \frac{\lambda^2}{2} [\qf(|\psi\rangle\langle\psi|)]_{kl}  +  (E_1 + E_2)/2,
	\end{equation}
	with the error term $E := (E_1 +  E_2)/2$
	which we can upper bound as
	\begin{equation}\label{errorterms}
		|E|		\leq		 |E_1|/2 +  |E_2|/2
		\leq   a +  b \lambda_2 .
	\end{equation}
\noindent \textbf{Global depolarising noise}\\
	In case global depolarising noise as $\rho := \lambda |\psi \rangle \langle \psi | + \tfrac{1-\lambda}{d} \mathrm{Id}$,
    we obtain the expression
	\begin{align*}
		\tr[(\partial_k \rho)(\partial_l \rho)]
		=  \frac{\lambda^2}{2} [\qf(|\psi\rangle\langle\psi|)]_{kl} +  E_{depol},
	\end{align*}
	where the error term is upper bounded by substituting $a=0$ and $\lambda_2 = (1-\lambda)/d$
	into Eq.~\eqref{errorterms} following Definition~\ref{definition:bterm} as
	\begin{equation*}
		E_{depol} \leq  b(1-\lambda)/d.
	\end{equation*}

In case of global depolarising noise we can also approximate the
Fisher information of the density matrix $\rho$ instead of the dominant eigenvector.
Using Lemmas~\ref{lemma:depol} and \ref{lemma_errorterm} we can obtain the approximation as
	\begin{align*}
	\tr[(\partial_k \rho)(\partial_l \rho)]
	=  \frac{\lambda}{2} [\qf(\rho)]_{kl}  +   (E_{1,depol} +E_{2})/2
\end{align*}
where we denote its error term $E'_{depol} := (E_{1,depol} +E_{2})/2$ and $|E'_{depol}| \leq |E_{1,depol}|/2 + |E_{2}|/2$ 
and under global depolarising noise we can substitute $\lambda_2 = (1-\lambda)/d$ into
the bound as $|E_2| \leq b (1+\lambda)\lambda_2$. We finally collect all terms and obtain the upper bound as
	\begin{equation*}
	|E'_{depol}| \leq  b [	\tfrac{(1-\lambda)^2}{d^2} +  \lambda \tfrac{1-\lambda}{d} ]/2 +  b(1-\lambda)(1+\lambda)/(2d)
	\in \mathcal{O}[(1-\lambda)/d].
\end{equation*}

\end{proof}

\section{Numerical Simulations}

\begin{figure}[tb]
	\begin{centering}
		\includegraphics[width=0.45\textwidth]{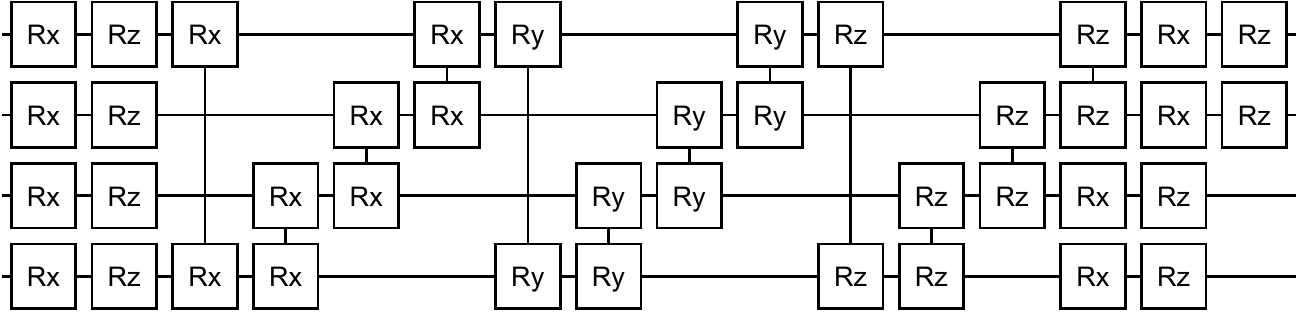}
		\caption{
				Example of a $4$-qubit ansatz circuit used for finding the ground state energy
				of the Hamiltonian in Eq.~\eqref{spin_ring}. It consists of single qubit $X$ and $Z$
				rotations and nearest neighbour $XX$ gates (which can be transformed into $YY$ and $ZZ$ gates by single qubit rotations).
			\label{circ}
		}
	\end{centering}
\end{figure}

\subsection{Simulations in Fig~\ref{fig:vqe_plots} \label{app:figvqeplots}}

We considered a hardware-efficient ansatz built of single qubit rotations and two-qubit $XX$ evolution gates, similar to
the circuit illustrated in Fig~\ref{circ}.
We simulated a physically motivated error model whereby single qubit gates undergo dephasing noise with 
probability $0.2 p_{err}$ to account for $T2$ relaxation and similarly undergo damping noise with a damping rate $0.01 p_{err}$
	to accuount for $T1$ relaxation, whereas two-qubit gates undergo 
two-qubit depolarisation with probability $p_{err} = 10^{-3}$. This error model and error severity is comparable to
state-of-the-art experimental hardware.

For the LiH simulation we used 5 ansatz layers and initialised the optimisation at random parameters around the
Hartree-Fock solution (HF parameters were disturbed with random numbers $-\pi \leq \Delta \theta_k \leq \pi$).
For the spin-ring simulations we used 8 ansatz layers and initialised the optimisation at random parameters
(note that barren plateaus are not necessarily present for local Hamiltonians~\cite{cerezo2020cost}).

At every iteration we compute the gradient vector $\underline{g}$ and the QFI matrix $\qf$.
We assume the gradient entries are estimated with a quantum computer via the parameter-shift rule,
i.e., as a difference of two energy evaluations each of which are of the form $\tr[\rho(\underline{\theta})  \mathcal{H}]$.
As such, we assume we have access to two copies of the quantum state and we have the ability to estimate the
error mitigated expected values as $\tr[\rho^2(\underline{\theta})  \mathcal{H}] / \tr[\rho^2(\underline{\theta})]$ via
the ESD/VD approach~\cite{koczor2020exponential, huggins2020virtual}.
Using the same quantum resources as for ESD/VD, we can also estimate the scalar products $\tr[\rho(\underline{\theta}_a) \rho(\underline{\theta}_b) ] $
for pairs of parameter vectors $\underline{\theta}_a$ and $\underline{\theta}_b$ which allows us to estimate the metric tensor in Result~\ref{result3} via parameter shifts~\cite{koczor2022quantum}.
We apply the simple regularisation from ref.~\cite{van2020measurement} as $[\qf + \eta \mathrm{Id}]^{-1}$ to
mitigate the ill-conditioned QFI matrix with a regularisation parameter $0.01$.
We choose the step size $\Delta t$ for each optimisation technique according to a linesearch: 
we test an increasing set of values $\Delta t$ and accept the largest value that does not increase the energy -- 
given determining the energy is significantly cheaper than determining the gradient vector and the metric tensor.

Let us now estimate the sampling cost of a single natural gradient iteration in our specific examples.
Generally, given a Hamiltonian $\mathcal{H} = \sum_{i=1}^{r_H} P_i c_i$  as a sum of Pauli strings $P_i$ with prefactors
$c_i$, the number of measurements to determine its expected value to precision $\epsilon$ is $N_E= T^2 / \epsilon^2$
with optimally distributed samples where each Pauli term is determined from $N_i = T |c_i|\sqrt{\mathrm{Var}[P_i] } / \epsilon^2 $ measurements.
A single gradient entry $g_k$ can thus be determined with parameter
shifts using on the order of $\propto 2 T^2 / \epsilon^2$ measurements where $T:= \sum_{i=1}^{r_H}  |c_i| \sqrt{ \mathrm{Var}[P_i] }$
and we find $T \propto 5.6$ for our LiH Hamiltonian. As such, for our LiH Hamiltonian, determining all $\nu = 103$ gradient entries to a typical precision $\epsilon = 0.05$ in practice requires on the order of $N_G \propto  2 \nu T^2 / \epsilon^2 \approx 2.6 \times 10^6$ measurements.
In contrast, the quantum Fisher information is independent of the Hamiltonian and determining all independent
entries to the same precision $\epsilon = 0.05$ requires $N_F \propto \nu (\nu+1)/(2 \epsilon^2) \approx 2.1 \times 10^6$
measurements.

In Fig.~\ref{fig:vqe_plots} we plot the energy deviation
from the exact ground state $E_{GS}$ that one would obtain from a noiseless device as $\langle \psi(\underline{\theta})  | \mathcal{H}| \psi(\underline{\theta}) \rangle -  E_{GS}$, but for the parameter values $\underline{\theta}$ we obtain from our noisy optimiser.

\subsection{Simulations in Fig~\ref{approxsimul}}

	In Fig~\ref{approxsimul} we assume 
	our aim is to find the ground state of the spin-ring Hamiltonian
	in Eq.~\eqref{spin_ring} using the noisy ansatz circuit shown in Fig.~\ref{circ}. This ansatz is composed of
	single-qubit $X$ and $Z$ rotations and two-qubit $XX$ gates (which can be transformed into $YY$ and $ZZ$ gates by single qubit rotations).
	We assume that every two-qubit gate undergoes two-qubit depolarisation 
	with a probability 
	$p_{err}$ whereas single-qubit gates undergo depolarisation with $0.1 p_{error}$
	and we also assume that error rates of the gates $p_{err}$
	slightly depend on the absolute values of gate parameters as $|\theta_k|$.

	We started the evolution from randomly chosen
	initial points $\underline{\theta}(0)$ in the vicinity of the optimal parameters $\underline{\theta}_{\mathrm{opt}}$
	that locally minimise the energy
	$E_{\mathrm{opt}} =\tr[ \rho(\underline{\theta}_{\mathrm{opt}}) \mathcal{H} ]$ to
	verify that our approach converges to the true optimum faster and more reliably than other techniques.
	Parameters of the ansatz circuit were evolved for a fixed number of $60$ steps using a
	step size $\kappa = 4 \, \Delta t = 0.2$  (this step size is slightly below the
	largest stable one) -- for the exact QFI simulation we used the same step size and
	evolved for $30$ iterations to account for the fact that its effective step size is at least by a factor of 2 larger
	via Result~\ref{result3}
	(and given extensive cost of classically computing the QFI matrix).
	We plot the energy distance $\Delta E = \tr[ \rho(\underline{\theta}) \mathcal{H} ] -E_{\mathrm{opt}} $ 
	at the parameters $\underline{\theta}$ we obtain from the noisy optimiser.

\end{document}